\documentclass[reqno, 11pt, a4paper]{amsart}
\usepackage[dvipsnames]{xcolor}
\usepackage[utf8]{inputenc}
\usepackage[T1]{fontenc}
\usepackage{lmodern}
\usepackage{amsmath}
\usepackage{amsfonts,amssymb}
\usepackage{amsthm}
\usepackage{enumerate}
\usepackage{booktabs}
\usepackage{tikz}
\usetikzlibrary{arrows}
\usetikzlibrary{arrows.meta,decorations.markings}
\usepackage{pgfplots}
\pgfplotsset{compat=1.18}
\tikzset{
  lbbv/.style={circle, fill=black, draw=black, inner sep=0pt, minimum size=4.5pt},
  lbwv/.style={circle, fill=white, draw=black, line width=0.6pt, inner sep=0pt, minimum size=4.5pt},
  lbc1/.style={draw=red!80!black, line width=0.8pt},
  lbc2/.style={draw=blue!75!black, line width=0.8pt},
  lbc3/.style={draw=green!50!black, line width=0.8pt},
  lbmid/.style={postaction={decorate, decoration={markings,
      mark=at position 0.5 with {\arrow[solid, xshift=1mm]{Stealth[length=2mm, width=1.7mm]}}}}},
  lbprop/.style={draw=black, line width=0.8pt, dash pattern=on 3pt off 2pt},
  lbwalk/.style={draw=black, line width=0.8pt, dash pattern=on 0.7pt off 2pt},
  lbbox/.style={rounded corners=6pt, draw=black, line width=0.5pt},
  lblab/.style={font=\footnotesize, text height=1.5ex, text depth=0.4ex, inner sep=2pt},
  lbelab/.style={font=\scriptsize, text height=1.3ex, text depth=0.3ex, inner sep=3pt}
}
\usepackage[includeheadfoot,margin=2cm]{geometry}
\usepackage[font=small,labelfont=bf,tableposition=top]{caption}
\usepackage{subcaption}
\definecolor{azulESI}{HTML}{1266AE}
\usepackage[bookmarks=false]{hyperref}
\hypersetup{
         colorlinks    =  true,
         citecolor    = azulESI,
         urlcolor  =  azulESI,
         linkcolor = azulESI
}

\numberwithin{equation}{section}
\newcommand{\tocappendixsingle}[3]{\indentlabel{\appendixname.\quad}#3}

\newtheoremstyle{mytheoremstyle} 
    {10pt}                   
    {8pt}                    
    {\itshape}               
    {}                       
    {\scshape}               
    {.}                      
    {.5em}                   
    {}                       
\theoremstyle{mytheoremstyle}
\newtheorem{theorem}{Theorem}[section]
\newtheorem{conjecture}[theorem]{Conjecture}
\newtheorem{corollary}[theorem]{Corollary}
\newtheorem{lemma}[theorem]{Lemma}
\newtheorem{proposition}[theorem]{Proposition}

\newtheoremstyle{definition} 
    {8pt}                    
    {5pt}                    
    {}                       
    {}                       
    {\scshape}               
    {.}                      
    {.5em}                   
    {}                       
\theoremstyle{definition}
\newtheorem{definition}[theorem]{Definition}
\newtheorem{example}[theorem]{Example}
\newtheorem{remark}[theorem]{Remark}

\DeclareMathOperator{\id}{id}
\DeclareMathOperator{\Tr}{Tr}

\tikzset{
  col1/.style={draw=red!80!black, thick},
  col2/.style={draw=blue!75!black, thick},
  col3/.style={draw=green!50!black, thick},
  bv/.style={circle, fill=black, draw=black, inner sep=0pt, minimum size=5pt},
  wv/.style={circle, fill=white, draw=black, inner sep=0pt, minimum size=5pt}
}
\title[Radially reducing random tensors]{Radially reducing random tensors for non-melonic cumulants and melonic factorization}
\author{Brayden Smith}
\address{McGill University, 805 Sherbrooke St W, Montreal, Quebec, H3A 2A7 Canada}
\email{brayden.smith@mail.mcgill.ca}

\begin{document}
\begin{abstract}
We study $U(N)^D$-invariant random tensors of rank $D\geq3$, given by a Gaussian measure perturbed by interactions. At large $N$ the dominant graphs are the melonic ones, a tree-like family that plays the role of planar maps for matrices. We prove two results. First, for the Gaussian measure, we find the leading large-$N$ term of the joint cumulant of any connected invariant, melonic or not, with any number of connected melonic invariants. The melonic invariants enter only through their numbers of vertices, as if each were a power of the squared norm of the tensor. The leading coefficient is the number of dominant Wick pairings of the first invariant times an explicit polynomial in these numbers. Second, order by order and for arbitrary interactions and scalings, if each Taylor coefficient of the two-point function stays bounded in $N$, then at leading order the expectations of products of melonic invariants are powers of the two-point function, so melonic invariants factorize. This applies to enhanced models whose large-$N$ limit is not melonic. Both proofs rest on a radial reduction: each melonic invariant is replaced by a power of the squared norm, which for a Gaussian tensor is a Gamma variable independent of the direction of the tensor. An appendix treats real tensors with $O(N)^D$ symmetry, including the tetrahedral model.
\end{abstract}
 \maketitle%
{\small\setcounter{tocdepth}{1}\tableofcontents}
\section{Introduction}\label{sec:intro}

Random tensor models generalize random matrix models~\cite{ambjorn1991three,sasakura1991tensor,gross1992tensor}. A random tensor $T_{a_1\cdots a_D}$ of rank $D\ge3$, with a law invariant under $U(N)^D$, is studied through the expectations of its invariants, as a random matrix is through its moments $\Tr(M^k)$. Where $\Tr(M^k)$ corresponds to a cycle, an invariant $\Tr_B(T,\bar T)$ corresponds to a bipartite $D$-coloured graph $B$, with a vertex for each $T$ and each $\bar T$ and an edge of colour $c$ for each contraction of $c$-th indices. These invariants are counted, through branched covers of the sphere, in~\cite{bengeloun2014counting}. Perturbing the Gaussian measure by invariant interactions turns these expectations into series in the couplings, whose coefficients are Gaussian integrals, computed by Wick's theorem as sums over coloured graphs weighted by powers of $N$~\cite{gurau2017random,gurau2012review,gurau2016invitation}.

The large-$N$ limit is governed by the \emph{melonic} graphs, obtained from the two-vertex melon by repeatedly inserting pairs of vertices joined by $D-1$ edges. They play the role of planar maps~\cite{thooft1974planar,brezin1978planar}, dominate the $1/N$ expansion~\cite{gurau2011expansion,gurau2011arbitrary,gurau2012complete,bonzom2011critical} of the coloured models of~\cite{gurau2011colored} and of uncoloured models~\cite{uncolouring}, and are geometrically branched polymers~\cite{gurau2014melons}. A melonic model has action $N^{D-1}\bigl[T\cdot\bar T+\sum_ig_i\Tr_{X_i}(T,\bar T)\bigr]$ with every interaction $X_i$ melonic, and its full two-point function $x$ solves the melonic equation $x+\sum_ip_ig_ix^{p_i}=1$, where $X_i$ has $2p_i$ vertices~\cite[(3.21)--(3.23)]{uncolouring}. Unlike the planar limit of matrix models, this limit is expected to be Gaussian: at large $N$ a melonic model should behave like a Gaussian tensor of covariance $x$~\cite{uncolouring,bonzom2013revisiting}. This paper makes this statement exact coefficient by coefficient in the couplings, in two directions: for general observables in melonic models, and for melonic observables in general models.

\subsection*{Cumulants}
Gur\u{a}u's universality theorem~\cite[Thm.~2]{gurau2014universality} is the rigorous form of the Gaussian limit. It shows that a trace-invariant measure whose cumulants obey a uniform bound in $N$, a \emph{properly uniformly bounded} measure, converges in distribution to a Gaussian tensor. The hypothesis has been established only for quartic interactions at small positive coupling, by the loop vertex expansion~\cite{rivasseau2007constructive,magnen2008constructive}, applied to tensors in~\cite{Gur13,delepouve2014universality} (see also \cite[App.~A.2]{gurau2014universality} and the review~\cite{rivasseau2016constructive}). The result of~\cite{delepouve2014universality} covers any $n_Q$ quartic interactions with a common coupling $0\le\lambda<1/(8n_Q)$. Beyond quartic order, Borel--Le~Roy summability in the coupling is known~\cite{LR16}, but not with the dependence on $N$ that a large-$N$ limit needs, and the uniform constructive results for interactions of higher order concern matrices~\cite{KRS17}.

Coefficient by coefficient the question is combinatorial. The coefficient of $g_{i_1}\cdots g_{i_q}$ in $\mathbb{E}_g[\Tr_B]$ is, up to a sign and a factorial, the joint cumulant $\kappa(\Tr_B,\Tr_{X_{i_1}},\dots,\Tr_{X_{i_q}})$ under the Gaussian measure, a finite sum over connected Feynman graphs. The coefficientwise content of universality is therefore the leading order in $N$ of these Gaussian cumulants. Gur\u{a}u bounds them graph by graph~\cite{gurau2014universality}, but the leading coefficients are not known in general.

Pagliaroli, P\'erez-S\'anchez and Smith~\cite{BTM} introduced positivity bootstraps for tensor integrals and, from series computed with \texttt{feyntensor}~\cite{feyntensor}, formulated two conjectures on these cumulants in rank three. The present author is a coauthor of~\cite{BTM}; the conjectures are due to Pagliaroli and P\'erez-S\'anchez.

\begin{conjecture}[{\cite[Conj.~I]{BTM}}]\label{conj:I}
Let $B$ be a connected $3$-coloured graph and let
\begin{equation*}
I_{i_1,\dots,i_q;N}[B]=N^{2q-1}\int^{\mathrm{conn}}\Tr_B(T,\bar T)\cdot P_{i_1}\cdots P_{i_q}\,d\mu_N(T,\bar T),\qquad i_1,\dots,i_q\in\{1,2,3\}.
\end{equation*}
Then the leading order in $N$ of $I_{i_1,\dots,i_q;N}[B]$ depends on the colours $\{i_1,\dots,i_q\}$ only through their number $q$.
\end{conjecture}

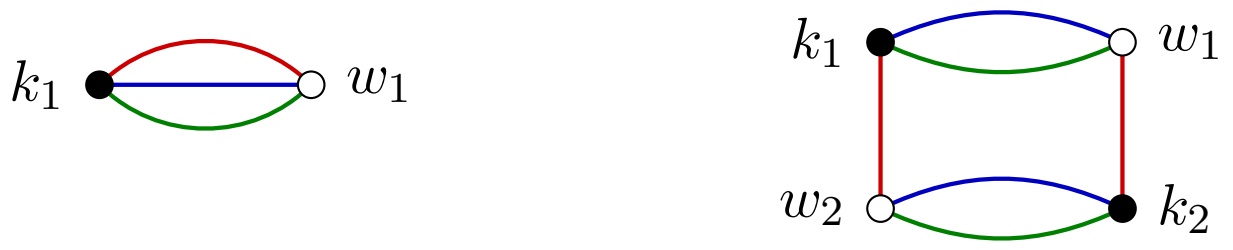
\begin{figure}[ht]
\centering
\begin{subfigure}[b]{0.3\textwidth}
\centering
\begin{tikzpicture}
\coordinate (k) at (0,0); \coordinate (w) at (1.4,0);
\draw[col1] (k) to[bend left=45] (w);
\draw[col2] (k) -- (w);
\draw[col3] (k) to[bend right=45] (w);
\node[bv, label=left:$k_1$] at (k) {};
\node[wv, label=right:$w_1$] at (w) {};
\path (0,-1.1) -- (0,1.1);
\end{tikzpicture}
\caption{melon $T\cdot\bar T$}
\end{subfigure}
\begin{subfigure}[b]{0.3\textwidth}
\centering
\begin{tikzpicture}
\coordinate (k1) at (0,1.1); \coordinate (w1) at (1.6,1.1);
\coordinate (w2) at (0,0); \coordinate (k2) at (1.6,0);
\draw[col2] (k1) to[bend left=25] (w1);
\draw[col3] (k1) to[bend right=25] (w1);
\draw[col2] (w2) to[bend left=25] (k2);
\draw[col3] (w2) to[bend right=25] (k2);
\draw[col1] (k1) -- (w2);
\draw[col1] (w1) -- (k2);
\node[bv, label=left:$k_1$] at (k1) {};
\node[wv, label=right:$w_1$] at (w1) {};
\node[wv, label=left:$w_2$] at (w2) {};
\node[bv, label=right:$k_2$] at (k2) {};
\end{tikzpicture}
\caption{pillow $P_1$}
\end{subfigure}
\begin{subfigure}[b]{0.3\textwidth}
\centering
\begin{tikzpicture}
\coordinate (k1) at (0,1.1); \coordinate (w1) at (1.6,1.1);
\coordinate (w2) at (1.6,0); \coordinate (k2) at (0,0);
\draw[col1] (k1) to[bend left=25] (w1);
\draw[col2] (k1) -- (w1);
\draw[col3] (k1) to[bend right=25] (w1);
\draw[col1] (k2) to[bend left=25] (w2);
\draw[col2] (k2) -- (w2);
\draw[col3] (k2) to[bend right=25] (w2);
\node[bv, label=left:$k_1$] at (k1) {};
\node[wv, label=right:$w_1$] at (w1) {};
\node[wv, label=right:$w_2$] at (w2) {};
\node[bv, label=left:$k_2$] at (k2) {};
\end{tikzpicture}
\caption{double melon $\mathcal{M}_2$}
\end{subfigure}
\caption{Rank three. Colour $1$ is red, colour $2$ blue and colour $3$ green; black vertices carry $\bar T$ and white vertices carry $T$. (a) The melon: one black and one white vertex joined by all $D$ colours. (b) The pillow $P_c$ for $c=1$. Its two \emph{dipoles} $(k_1,w_1)$ and $(k_2,w_2)$ are joined by the two colours $\neq c$, and the colour-$c$ edges $k_1w_2$ and $k_2w_1$ join them. (c) The double melon $\mathcal{M}_2$ has the same four vertices, with each dipole replaced by a melon.}\label{fig:graphs}
\end{figure}

Here $P_c$ is the pillow whose colour-$c$ edges cross between its two dipoles (Fig.~\ref{fig:graphs}), $d\mu_N$ is the Gaussian measure with covariance $N^{-2}$, and $\int^{\mathrm{conn}}$ keeps connected Feynman graphs only; $B$ may be non-melonic and of any genus. Expanding $\mathbb{E}_g[\Tr_B]$ in the quartic model with action $N^2\bigl[T\cdot\bar T+\frac g2(P_1+P_2+P_3)\bigr]$, the coefficient of $g^q$ is
\begin{equation*}
\frac{(-1)^q}{2^q\,q!}\,N\sum_{i_1,\dots,i_q=1}^{3}I_{i_1,\dots,i_q;N}[B]
\end{equation*}
(see the remark after the proof of Corollary~\ref{cor:moments}). Conjecture~\ref{conj:I} thus asserts that the $3^q$ summands agree at leading order.

Conjecture~II~\cite[(5.7)]{BTM} concerns connected melonic graphs $C_1,\dots,C_q$. It states that the limit of $N^{2q-3}\int^{\mathrm{conn}}\Tr_{C_1}\cdots\Tr_{C_q}\,d\mu_N$ depends only on the numbers of vertices of the $C_i$. P\'erez-S\'anchez proved Conjecture~II~\cite[Thm.~4.2]{TwofoldUniversality}, conjectured that the limit is given by explicit melonic polynomials, and verified this for a finite list of vertex numbers~\cite[Thm.-Conj.~4.4]{TwofoldUniversality}. Conjecture~\ref{conj:I} for melonic $B$ is the case of Conjecture~II in which all graphs but one are pillows.

The leading order of $I_{i_1,\dots,i_q;N}[B]$ for non-melonic $B$ lies below $N^0$, beyond the reach of Conjecture~II. Two questions are thus open: whether the leading order of these cumulants depends only on the numbers of vertices of the interactions, also for non-melonic $B$ and for interactions of any order, and what its value is.

\subsection*{Factorization}
In random matrix models, expectations of products of traces factorize at large $N$ into products of expectations~\cite{thooft1974planar,voiculescu1991limit}. For tensors, Gur\u{a}u showed that under a properly uniformly bounded measure the expectation of a connected melonic observable with $2n$ vertices converges, after division by $N$, to the $n$-th power of the limiting covariance~\cite[\S4.5.1, (64)]{gurau2014universality}. Bonzom, Gur\u{a}u and Rivasseau~\cite[(3.19)]{uncolouring} argued for the same law in interacting models, with the full two-point function in place of the covariance and with non-melonic observables vanishing at that scale. Bonzom derived this single-trace law, $N^{-1}\langle\Tr_B\rangle=G^{|B|/2}$ for melonic $B$ with $G$ the two-point function~\mbox{\cite[(6)]{bonzom2013revisiting}}, from the Schwinger--Dyson equations of tensor models~\cite{gurau2011virasoro,gurau2012schwinger}, taking melonic dominance~\mbox{\cite[(5)]{bonzom2013revisiting}} and the factorization of multi-trace expectations~\mbox{\cite[(8)]{bonzom2013revisiting}} as assumptions. For the quartic melonic model, the correlators of its intermediate-field matrix model satisfy, once the leading order is removed, the blobbed topological recursion~\cite{bonzom2018blobbed}.

Factorization of all multi-trace expectations had been conjectured for Gaussian complex tensors~\cite[Conj.~5.1 and \S4.3.3]{CGL24}. Gur\u{a}u, Joos and Sudakov showed that it fails~\cite{gurau2025nonfactor}: for Gaussian real tensors of rank $D\ge3$, the cumulant of a product of trace invariants is not always suppressed relative to the product of their expectations. Their argument is statistical and shows that most large graphs fail; they also show that factorization holds on the melonic family. Berthold and Keppler later gave the first explicit counterexamples for real Gaussian tensors, the smallest with $16$ vertices~\cite{BK26}. Carrozza, Chevrier and Lionni~\cite{CCL26} give an explicit non-factorizing invariant for Gaussian and Haar-distributed complex tensors. They prove factorization for every family of trace invariants with tree-like dominant pairings, which includes the melonic graphs and the connected planar $3$-coloured graphs~\cite[Thm.~4.7 and Thm.~2.13]{CCL26}, the latter case resting on~\cite{bonzom2018balls}. In tensor models, then, factorization holds only on restricted families of observables, and the melonic family is the basic one.

These results concern the Gaussian or Haar measure. Interacting models need not even have the melonic scaling: Bonzom, Delepouve and Rivasseau~\cite{bonzom2015enhancing} showed that enhancing non-melonic quartic interactions yields different large-$N$ limits, which mix melonic and planar structures (see~\cite{bonzom2016large} for a review of such universality classes). For a family of such models they proved that the expectation of a single tree of necklaces, melonic graphs included, factorizes over its necklaces~\cite[Thm.~3]{bonzom2015enhancing}, but left the full model, with necklaces of three colour types, unsolved. By contrast, the totally unbalanced generalized melonic interactions, which include non-melonic ones, still have a Gaussian large-$N$ limit~\cite{bonzom2019generalized}. P\'erez-S\'anchez showed that in an arbitrary model the large-$N$ melonic moments, when they converge, depend on the observable only through its number of vertices~\cite{RedundantMoments}. Whether melonic multi-traces factorize for interacting measures with arbitrary interactions and scalings, and what the common value of the melonic moments is, has not been settled.

\subsection*{Main results}
Fix $D\geq 3$. Melonic graphs are defined in Definition~\ref{def:melonic}; the pillows are exactly the melonic graphs with four vertices. The joint cumulant $\kappa(Y_0,\dots,Y_q)$ of random variables $Y_0,\dots,Y_q$ with finite moments is defined by the moment--cumulant formula
\begin{equation*}
\mathbb{E}[Y_0\cdots Y_q]=\sum_{\mathcal{P}}\prod_{V\in\mathcal{P}}\kappa(Y_j:j\in V),
\end{equation*}
the sum running over partitions $\mathcal{P}$ of $\{0,\dots,q\}$. The connected integral $\int^{\mathrm{conn}}$ of \cite{BTM} is a joint cumulant (Section~\ref{ssec:gaussian}). Given a connected $D$-coloured graph $B$ with $n$ black vertices, we write $\psi_0(B)$ and $\ell(B)$ for the maximal exponent and the number of maximizing Wick pairings (Section~\ref{ssec:gaussian}) in the Gaussian expectation
\begin{equation*}
N^{-1}\mathbb{E}[\Tr_B]=\ell(B)\,N^{\psi_0(B)-1}+O\bigl(N^{\psi_0(B)-2}\bigr).
\end{equation*}
If $B$ is melonic, then $\psi_0(B)=\ell(B)=1$. In Gur\u{a}u's terms, $1-\psi_0(B)$ is the convergence order of $B$ and $\ell(B)$ its number of minimal covering graphs (Remark~\ref{rem:gurau}). Let
\begin{equation}\label{eq:Phi}
\mathrm{Mel}_{q+1}(p_0,p_1,\dots,p_q)=p_0p_1\cdots p_q\,\frac{(S-1)!}{(S-q)!},\qquad S=p_0+p_1+\dots+p_q,
\end{equation}
be the melonic polynomials of P\'erez-S\'anchez~\cite[Def.~4.3]{TwofoldUniversality}, with $\mathrm{Mel}_1(p_0)=1$; they are symmetric in their arguments. For pillows ($p_1=\dots=p_q=2$) and $p_0=n$, $\mathrm{Mel}_{q+1}(n,2,\dots,2)=c_{n,q}=2^q q!\,\frac{n}{2q+n}\binom{2q+n}{q}$.

\begin{theorem}\label{thm:main}
Let $D\geq 3$, let $B$ be a connected $D$-coloured graph with $2n$ vertices, and let $X_1,\dots,X_q$ be connected melonic $D$-coloured graphs with $2p_1,\dots,2p_q$ vertices. Then
\begin{align*}
\mathcal{I}_N[B;X_1,\dots,X_q]&:=N^{(D-1)q-1}\,\kappa\bigl(\Tr_B,\Tr_{X_1},\dots,\Tr_{X_q}\bigr)\\
&\phantom{:}=\ell(B)\,\mathrm{Mel}_{q+1}(n,p_1,\dots,p_q)\,N^{\psi_0(B)-1}+O\bigl(N^{\psi_0(B)-2}\bigr).
\end{align*}
In particular the leading term has degree exactly $\psi_0(B)-1$ and depends on the $X_j$ only through their numbers of vertices.
\end{theorem}

That is, at leading order each melonic interaction with $2p$ vertices behaves like $(T\cdot\bar T)^p/N^{p-1}$. In rank three we have
\begin{equation*}
\mathcal{I}_N[B;P_{i_1},\dots,P_{i_q}]=I_{i_1,\dots,i_q;N}[B],
\end{equation*}
so Theorem~\ref{thm:main} proves Conjecture~\ref{conj:I} for all $B$ and gives the leading term $\ell(B)\,c_{n,q}\,N^{\psi_0(B)-1}$. For the melon this gives $2^qq!\,\mathrm{Cat}_q=2,16,240,5376,161280,\dots$, the boldface terms of \cite[\S5 and App.~C]{BTM}.

\paragraph*{Example.} Take $D=3$ and $B=K_{3,3}$ (Fig.~\ref{fig:invariants}(b)), which is non-melonic, with $n=3$, $\psi_0(B)=0$ and $\ell(B)=3$ (Example~\ref{ex:Bstar}). For one pillow, of any colour $c$, Theorem~\ref{thm:main} gives
\begin{equation*}
I_{c;N}[K_{3,3}]=3\,\mathrm{Mel}_2(3,2)\,N^{-1}+O(N^{-2})=\frac{18}{N}+O(N^{-2}),\qquad \mathrm{Mel}_2(3,2)=3\cdot2\cdot\frac{4!}{4!}=6.
\end{equation*}
For two pillows, of any colours $c,d$, it gives
\begin{equation*}
I_{c,d;N}[K_{3,3}]=3\,\mathrm{Mel}_3(3,2,2)\,N^{-1}+O(N^{-2})=\frac{216}{N}+O(N^{-2}),\qquad \mathrm{Mel}_3(3,2,2)=3\cdot2\cdot2\cdot\frac{6!}{5!}=72.
\end{equation*}

Unlike Conjecture~\ref{conj:I}, which is stated in rank three, Theorem~\ref{thm:main} holds in every rank $D\geq3$. Remark~\ref{rem:rank} discusses the restriction to rank three in \cite{BTM}. The fact that the leading order depends on different colourings only through their number holds for all melonic graphs, not just pillows. For non-melonic $B$ the rank enters the leading term only through the Gaussian data $\psi_0(B)$ and $\ell(B)$, while $\mathrm{Mel}_{q+1}$ does not depend on $D$.

Relative to \cite{BTM,TwofoldUniversality} the non-melonic case is new. Gur\u{a}u's universality theorem \cite{gurau2014universality} also covers all observables, but for the resummed measure under a boundedness hypothesis. Taking $B$ melonic gives a closed form for Conjecture~II in any rank.

\begin{corollary}\label{cor:melonic}
Let $C_0,\dots,C_q$ be connected melonic $D$-coloured graphs with $2p_0,\dots,2p_q$ vertices. Then
\begin{equation*}
\lim_{N\to\infty}N^{(D-1)q-1}\,\kappa\bigl(\Tr_{C_0},\dots,\Tr_{C_q}\bigr)=\mathrm{Mel}_{q+1}(p_0,\dots,p_q).
\end{equation*}
\end{corollary}

Corollary~\ref{cor:melonic} proves \cite[Thm.-Conj.~4.4]{TwofoldUniversality} for all vertex numbers and in every rank $D\geq 3$. P\'erez-S\'anchez had proved it for $q=0$, that is, for a single melonic graph, where the limit is $\mathrm{Mel}_1(p_0)=1$. He also verified it by exact computation in rank three for a finite list of vertex numbers, and announced a proof of the general case by another strategy~\cite{PSmelonic}. His proof of Conjecture~II \cite[Thm.~4.2]{TwofoldUniversality} gives the independence of the limit from the graphs and from the rank. It follows that two interacting melonic models whose couplings agree after summing over interactions of equal size have the same large-$N$ melonic cumulants \cite[Thm.~5.1]{TwofoldUniversality}.

Writing $J_{p_0,\dots,p_q}$ for this limit in rank three, which is $\lim_NJ_N(C_0,\dots,C_q)$ in the notation of \cite[(5.6)]{BTM}, Corollary~\ref{cor:melonic} reproduces all ten values $J_{2,3}=6$, $J_{3,3}=9$, $J_{2,2,3}=72$, $J_{2,3,3}=126$, $J_{3,3,3}=216$, $J_{2,2,2,3}=1344$, $J_{3,4}=12$, $J_{4,4}=16$, $J_{2,3,4}=192$, $J_{2,2,3,3}=2592$ computed in \cite[(5.9)]{BTM}. 

We can translate Theorem~\ref{thm:main} into the language of moments by summing over couplings. Let $X_1,\dots,X_L$ be connected melonic $D$-coloured graphs, $X_i$ having $2p_i$ vertices. For formal couplings $g=(g_1,\dots,g_L)$ set
\begin{equation*}
V_g=N^{D-1}\sum_{i=1}^L g_i\Tr_{X_i},\qquad \mathbb{E}_g[\mathcal{O}]=\frac{\mathbb{E}\bigl[\mathcal{O}\,e^{-V_g}\bigr]}{\mathbb{E}\bigl[e^{-V_g}\bigr]},
\end{equation*}
where numerator and denominator are expanded in powers of $g$ under the Gaussian measure, so that $\mathbb{E}_g[\mathcal{O}]$ is a formal power series in $g$ whose coefficients are Laurent polynomials in $N$. Let $x(g)$ be the unique formal power series in $g$ with
\begin{equation}\label{eq:twopoint}
x(g)+\sum_{i=1}^L p_ig_i\,x(g)^{p_i}=1,\qquad x(0)=1.
\end{equation}
For a multi-index $\alpha\in\mathbb{N}^L$ write $g^\alpha=\prod_ig_i^{\alpha_i}$, $|\alpha|=\sum_i\alpha_i$, $\alpha!=\prod_i\alpha_i!$, and $[g^\alpha]$ for the coefficient of $g^\alpha$ in a power series. Here and below, $a^{(\alpha)}$ denotes the list in which each entry $a_i$ of a list $a=(a_1,\dots,a_L)$ appears $\alpha_i$ times.

\begin{corollary}\label{cor:moments}
Let $B$ be a connected $D$-coloured graph with $2n$ vertices. Then, for every $\alpha\in\mathbb{N}^L$,
\begin{equation*}
[g^\alpha]\,\mathbb{E}_g[\Tr_B]=\ell(B)\bigl([g^\alpha]\,x(g)^n\bigr)N^{\psi_0(B)}+O\bigl(N^{\psi_0(B)-1}\bigr),\qquad [g^\alpha]\,x(g)^n=\frac{(-1)^{|\alpha|}}{\alpha!}\,\mathrm{Mel}_{|\alpha|+1}\bigl(n,p^{(\alpha)}\bigr).
\end{equation*}
In particular the leading coefficient is non-zero.
\end{corollary}

That is, for melonic models the leading order of the Taylor coefficients of $\mathbb{E}_g[\Tr_B]$ depends on the potential only through the number of vertices in each interaction.

The couplings $g_i$ are formal: no sign or smallness condition is imposed, and $\mathbb{E}_g[\Tr_B]$ is in general a divergent series. Corollary~\ref{cor:moments} is a statement about each Taylor coefficient separately, with a constant in the error term that depends on $\alpha$; by itself it says nothing about resumming the series at a fixed coupling. Only the sizes $p_i$ enter, and the $X_i$ need not be distinct.

Corollary~\ref{cor:moments} is thus a coefficientwise counterpart of Gur\u{a}u's second universality theorem \cite[Thm.~2]{gurau2014universality} for melonic models. Where Gur\u{a}u's theorem holds with Borel summability uniform in $N$, as for quartic interactions at small positive coupling \cite{delepouve2014universality}, the limits of $N^{-\psi_0(B)}\mathbb{E}_g[\Tr_B]$, namely $\ell(B)x(g)^n$, have the asymptotic expansion given term by term by Corollary~\ref{cor:moments}. This uses that uniform Borel summability gives bounds on the Taylor remainders that are uniform in $N$. Gur\u{a}u's bounds on Feynman amplitudes also hold graph by graph \cite{gurau2014universality}; what is new here is that the leading coefficients are identified exactly, for each tuple of interactions, with no boundedness hypothesis.

Consider for example the pillows $X_c=P_c$, $c\in[D]$, with couplings $g_c/2$. Then \eqref{eq:twopoint} reads $x+Gx^2=1$ with $G=\sum_cg_c$, so that, coefficientwise, $\lim_{N\to\infty}N^{-1}\mathbb{E}_g[T\cdot\bar T]=\bigl(\sqrt{1+4G}-1\bigr)/(2G)$, which equals $1$ at $G=0$. Equation~\eqref{eq:twopoint} is the melonic two-point equation, which goes back to the analysis of the melonic sector in \cite{bonzom2011critical,uncolouring}; in this normalization it is \cite[(7)]{bonzom2013revisiting}. It gives the two-point function of \cite{nguyen2015analysis,BTM} for the pillows. In the hexic cyclic model of \cite[\S4.2]{BTM} it reads $x+3t_6x^3=1$, as in \cite[after (58)]{lionni2019multi}, up to the sign of the coupling.

Theorem~\ref{thm:main} and Corollaries~\ref{cor:melonic} and~\ref{cor:moments} hold for real tensors with $O(N)^D$ symmetry \cite{carrozza2016on,klebanov2017uncolored}, whose invariants are indexed by $D$-coloured graphs that need not be bipartite (Theorem~\ref{thm:mainO} and Corollary~\ref{cor:melonicO}). The observable $B$ may then be non-bipartite, and $(\psi_0,\ell)$ become their real analogues $(\psi_0^O,\ell^O)$. The cumulants in Theorem~\ref{thm:main} and Corollary~\ref{cor:melonic} gain a factor $2^q$. With the potential $\tfrac12N^{D-1}\sum_ig_i\Tr_{X_i}$ this factor cancels in Corollary~\ref{cor:moments}, so real and complex melonic models have the same series $x(g)$. The proofs are those of the complex case, with one extra kind of Wick pairing (\hyperref[app:real]{Appendix}).

\subsection*{Melonic observables in arbitrary models}
Melonic observables keep this structure when the interactions are not melonic. Let $X_1,\dots,X_L$ be arbitrary connected $D$-coloured graphs, $X_i$ having $2n_i$ vertices, let $s_1,\dots,s_L\in\mathbb{R}$, and set
\begin{equation}\label{eq:Vs}
V_g=\sum_{i=1}^Lg_iN^{s_i}\Tr_{X_i},
\end{equation}
with $\mathbb{E}_g$ defined from $V_g$ as before Corollary~\ref{cor:moments}. Melonic models are the case of melonic $X_i$ and $s_i=D-1$. The scalings giving a non-trivial limit are not known for general interactions~\cite{RedundantMoments}. We assume only that the Taylor coefficients of the two-point function stay bounded.

\begin{theorem}\label{thm:arbitrary}
Suppose that for every $\alpha\in\mathbb{N}^L$ the coefficient $[g^\alpha]\,N^{-1}\mathbb{E}_g[T\cdot\bar T]$ is bounded as $N\to\infty$. Then it converges to $[g^\alpha]\,m_2(g)$ for a formal power series $m_2$ with $m_2(0)=1$, and for all connected melonic $C_1,\dots,C_r$ with $2p_1,\dots,2p_r$ vertices and every $\alpha$,
\begin{equation*}
\lim_{N\to\infty}[g^\alpha]\,N^{-r}\,\mathbb{E}_g\Bigl[\prod_{j=1}^r\Tr_{C_j}\Bigr]=[g^\alpha]\,m_2(g)^{p_1+\dots+p_r}.
\end{equation*}
\end{theorem}

Thus, whenever the hypothesis holds, every melonic moment is, coefficientwise, at leading order a power of the two-point function, and products of melonic traces factorize. The content of Theorem~\ref{thm:arbitrary} is a transfer: a bound on one observable, $T\cdot\bar T$, gives the limits of all melonic multi-traces.

Theorem~\ref{thm:arbitrary} gives, coefficientwise, both the factorization of products and the single-trace law $N^{-1}\mathbb{E}_g[\Tr_C]\to m_2^p$, for connected melonic $C$ with $2p$ vertices, of~\cite[(64)]{gurau2014universality}, \cite[(3.19)]{uncolouring} and~\cite{bonzom2013revisiting}, which for melonic observables are equivalent. For non-melonic observables they can differ (Section~\ref{sec:discussion}).

By the Ward identity, the hypothesis is equivalent to the bound $O(N^D)$ on every connected vacuum Feynman graph, that is, to the free energy $N^{-D}\log\mathbb{E}[e^{-V_g}]$ having bounded Taylor coefficients (Corollary~\ref{cor:freeenergy}). Since this is the usual form of a large-$N$ expansion, the hypothesis can be read off from any result establishing one. It holds, for instance, at the canonical scaling $s_i=D-1$ for arbitrary interactions, by the known bound on the degree \cite{gurau2011expansion,gurau2012complete,uncolouring,gurau2017random}. Beyond such results it is not verified intrinsically, since we have no criterion on the $(X_i,s_i)$ that guarantees it.

Theorem~\ref{thm:arbitrary} also holds for real tensors, with $T\cdot T$ in place of $T\cdot\bar T$ and interactions $X_i$ that need not be bipartite (Section~\ref{ssec:arbreal}). We check the hypothesis in three explicit models, each time by Theorem~\ref{thm:main} or by results from outside this paper. In melonic models, $m_2=x(g)$ (Corollary~\ref{cor:overlap}). The maximally enhanced quartic model in rank four of~\cite{bonzom2015enhancing} has a large-$N$ limit that mixes melonic and planar structures, and its Schwinger--Dyson equations are unsolved; its melonic sector, products included, is nevertheless determined by $m_2$ (Proposition~\ref{prop:enhanced}, using \cite[Thm.~1]{bonzom2015enhancing}). The third example is the real tetrahedral model of Carrozza and Tanasa~\cite{carrozza2016on}. A fermionic quantum-mechanical analogue is the model of Klebanov and Tarnopolsky~\cite{klebanov2017uncolored}, an uncoloured version of Witten's SYK-like tensor model~\cite{witten2019syk,gurau2017complete}. In the Carrozza--Tanasa model, melonic observables factorize to all orders, with $m_2=1+16g^2m_2^4$ (Proposition~\ref{prop:tetra}, using \cite[Prop.~2]{carrozza2016on}).

Along the way, the rescaled joint cumulants of the interactions on the unit sphere converge (Theorem~\ref{thm:factor}), and an exact Ward identity expresses the free energy and the interaction energies through $m_2$ (Proposition~\ref{prop:ward}). This identity is known for arbitrary interactions with degree-dependent scalings, which include melonic interactions at the canonical scaling~\cite{uncolouring,bonzom2013revisiting}; its proof applies to arbitrary scalings.

Theorem~\ref{thm:arbitrary} identifies the common value of the melonic moments of~\cite{RedundantMoments} as $m_2^p$ and, coefficientwise and whenever its hypothesis holds, extends the factorization of melonic multi-traces known for Gaussian real~\cite{gurau2025nonfactor} and Gaussian or Haar complex~\cite{CCL26} tensors to interacting measures with arbitrary interactions and scalings.

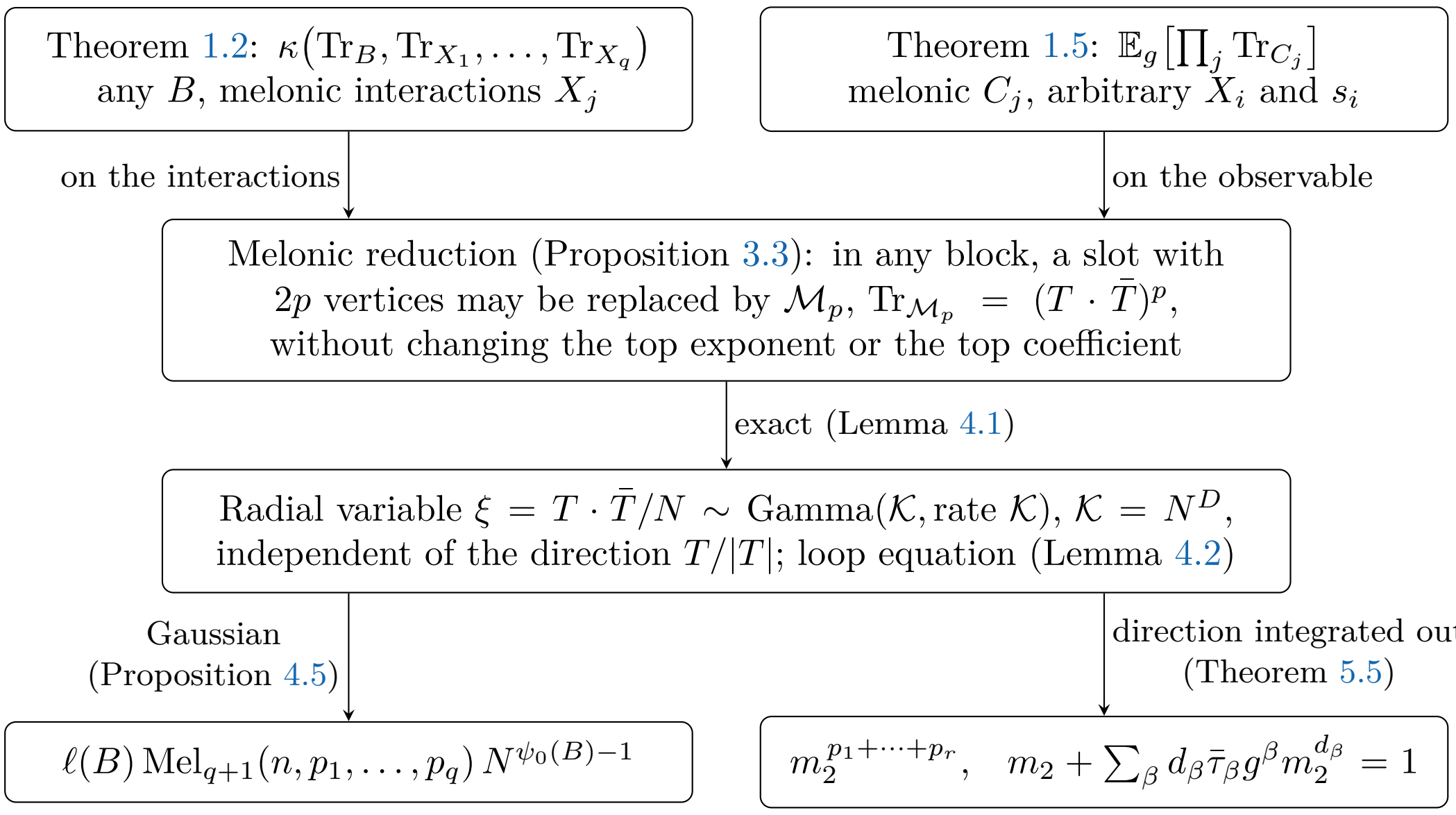
\begin{figure}[ht]
\centering
\begin{tikzpicture}[>=stealth,
  box/.style={draw, rounded corners=3pt, text width=6.2cm, align=center, inner sep=5pt, font=\small},
  wbox/.style={draw, rounded corners=3pt, text width=10.4cm, align=center, inner sep=5pt, font=\small},
  lab/.style={font=\footnotesize, align=center, inner sep=2pt}]
\node[box] (A) at (-3.6,0) {Theorem~\ref{thm:main}: $\kappa\bigl(\Tr_B,\Tr_{X_1},\dots,\Tr_{X_q}\bigr)$\\ any $B$, melonic interactions $X_j$};
\node[box] (B) at (3.6,0) {Theorem~\ref{thm:arbitrary}: $\mathbb{E}_g\bigl[\prod_j\Tr_{C_j}\bigr]$\\ melonic $C_j$, arbitrary $X_i$ and $s_i$};
\node[wbox] (R) at (0,-2.2) {Melonic reduction (Proposition~\ref{prop:melred}): in any block, a slot with $2p$ vertices may be replaced by $\mathcal{M}_p$, $\Tr_{\mathcal{M}_p}=(T\cdot\bar T)^p$, without changing the top exponent or the top coefficient};
\node[wbox] (G) at (0,-4.4) {Radial variable $\xi=T\cdot\bar T/N\sim\mathrm{Gamma}(\mathcal{K},\text{rate }\mathcal{K})$, $\mathcal{K}=N^D$, independent of the direction $T/|T|$; loop equation (Lemma~\ref{lem:loop})};
\node[box] (A2) at (-3.6,-6.6) {$\ell(B)\,\mathrm{Mel}_{q+1}(n,p_1,\dots,p_q)\,N^{\psi_0(B)-1}$};
\node[box] (B2) at (3.6,-6.6) {$m_2^{\,p_1+\dots+p_r}$,\quad $m_2+\sum_{\beta}d_\beta\bar\tau_\beta g^\beta m_2^{d_\beta}=1$};
\draw[->] (A.south) -- (A.south |- R.north) node[midway, lab, left] {on the interactions};
\draw[->] (B.south) -- (B.south |- R.north) node[midway, lab, right] {on the observable};
\draw[->] (R) -- (G) node[midway, lab, right] {exact (Lemma~\ref{lem:radial})};
\draw[->] (A2.north |- G.south) -- (A2.north) node[midway, lab, left] {Gaussian\\(Proposition~\ref{prop:gammalimit})};
\draw[->] (B2.north |- G.south) -- (B2.north) node[midway, lab, right] {direction integrated out\\(Theorem~\ref{thm:factor})};
\end{tikzpicture}
\caption{Architecture of the proofs. Both theorems reduce, by the same lemma, to cumulants of powers of $M=T\cdot\bar T$, which one loop equation computes: without interactions for Theorem~\ref{thm:main}, and with the direction of $T$ integrated out for Theorem~\ref{thm:arbitrary}. For real tensors $\xi\sim\mathrm{Gamma}(\mathcal{K}/2,\text{rate }\mathcal{K}/2)$. The equation for $m_2$ (lower right) re-expresses $m_2$ through the limits $\bar\tau_\beta$ of the sphere cumulants of the interactions; its content is that these limits exist.}\label{fig:architecture}
\end{figure}

\subsection*{Strategy}
Arbitrary observables in melonic models (Theorem~\ref{thm:main}) and melonic observables in arbitrary models (Theorem~\ref{thm:arbitrary}) rest on the same two ingredients (Fig.~\ref{fig:architecture}), each proved once. These proofs use only finite Wick combinatorics and a one-variable moment recursion.
\begin{enumerate}[(i)]
\item \emph{Melonic reduction} (Section~\ref{sec:reduction}). Splitting a $(D-1)$-dipole off a melonic graph turns it into a separate melon. A transposition argument shows that the pairings for which this changes the exponent are suppressed by $N^{-(D-2)}$. Iterating, a melonic graph with $2p$ vertices can be replaced by the disjoint union $\mathcal{M}_p$ of $p$ melons, in any block of a cumulant, without changing its top exponent or its top coefficient (Proposition~\ref{prop:melred}, Fig.~\ref{fig:cascade}). The argument needs no a priori bound and treats observables and interactions alike.
\item \emph{The radial variable} (Section~\ref{sec:radial}). Since $\Tr_{\mathcal{M}_p}=(T\cdot\bar T)^p$ and the radius $T\cdot\bar T$ is a Gamma variable independent of the direction of $T$, cumulants against products of melons factorize exactly into Gaussian data of $B$ and cumulants of one Gamma variable. The moment recursion of that variable, read as a loop equation, gives the melonic polynomials in the Gaussian case and, once the direction is integrated out, the two-point equation of an arbitrary model.
\end{enumerate}
Section~\ref{sec:proofs} assembles the proofs of the coefficientwise results, Section~\ref{sec:applications} treats melonic and enhanced models, and Section~\ref{sec:discussion} lists open questions. The \hyperref[app:real]{Appendix} treats real tensors.

\section{Setting}\label{sec:setting}

\subsection{Coloured graphs and invariants}\label{ssec:coloured}
Fix $D\geq 3$. A $D$-coloured graph $B$ on $n$ black vertices $\{1^\bullet,\dots,n^\bullet\}$ and $n$ white vertices $\{1^\circ,\dots,n^\circ\}$ is encoded by $(b_1,\dots,b_D)\in\mathfrak{S}_n^D$: the colour-$c$ edges join $j^\bullet$ to $b_c(j)^\circ$. We attach an index $a_j^c\in[N]$ to the colour-$c$ edge at $j^\bullet$. The associated $U(N)^D$-invariant is
\begin{equation*}
\Tr_B(T,\bar T)=\sum_a\prod_{j=1}^n\bar T_{a_j^1\cdots a_j^D}\prod_{i=1}^nT_{a^1_{b_1^{-1}(i)}\cdots a^D_{b_D^{-1}(i)}},
\end{equation*}
where the sum runs over the $N^{Dn}$ assignments $a=(a_j^c)_{j\in[n],\,c\in[D]}$. It is homogeneous of degree $n$ in $T$ and in $\bar T$ and factorizes over connected components. The melon $\ominus$ is the graph with $n=1$, one black and one white vertex joined by $D$ edges; its invariant is $T\cdot\bar T$.

\subsection{Melonic graphs}\label{ssec:melonic}
A \emph{$(D-1)$-dipole} of a $D$-coloured graph is a pair $\delta=(k,w)$ of a black and a white vertex joined by edges of all colours except one colour $c$. \emph{Inserting} a $(D-1)$-dipole on an edge of colour $c$ from $j^\bullet$ to $i^\circ$ means deleting that edge, adding a white vertex $w$ and a black vertex $k$ joined by edges of all colours except $c$, and adding colour-$c$ edges from $j^\bullet$ to $w$ and from $k$ to $i^\circ$.

\begin{definition}\label{def:melonic}
A $D$-coloured graph is \emph{melonic} if it can be obtained from the melon by a finite sequence of $(D-1)$-dipole insertions.
\end{definition}

A melonic graph with $2p$ vertices arises from $p-1$ insertions and is connected. The melonic graphs with four vertices are exactly the pillows $P_c$ ($c\in[D]$): one dipole is inserted on the colour-$c$ edge of the melon, so that the colour-$c$ edges cross between the two dipoles. In rank three, $P_1$ is $T_{a_1a_2a_3}\bar T_{b_1a_2a_3}T_{b_1b_2b_3}\bar T_{a_1b_2b_3}$, denoted $\square_1$ in \cite[(2.2a)]{BTM}. \emph{Removing} a $(D-1)$-dipole $\delta=(k,w)$ missing the colour $c$ from a graph $B$ means deleting $k$, $w$ and their edges, and joining the colour-$c$ neighbours $b_c^{-1}(w)^\bullet$ and $b_c(k)^\circ$ by a colour-$c$ edge. We write $B\setminus\delta$ for the result. We use two facts.
\begin{enumerate}
\item[(M1)] If $Y$ is melonic with at least four vertices and $\delta$ is the dipole created by the last insertion in a sequence producing $Y$, then $Y\setminus\delta$ is melonic: removing $\delta$ restores the edge on which it was inserted.
\item[(M2)] For $D\geq 3$, a connected graph $B$ is melonic if and only if its Gaussian expectation is of order $N$, that is $\psi_0(B)=1$, and then $\ell(B)=1$. Moreover, $\psi_0(B)\leq1$ for every connected $B$ \cite[\S4.3.1 and Lemma~4]{gurau2014universality}. Since the exponents are integers, $\psi_0(B)\leq0$ for every connected non-melonic $B$.
\end{enumerate}
In rank three, a connected $3$-coloured graph $B$ with $2n$ vertices is a ribbon graph, with cyclic order $(1,2,3)$ at black and $(3,2,1)$ at white vertices, whose faces are its bicoloured cycles. Its \emph{genus} $g(B)$ is given by $2-2g(B)=2n-3n+F_B$, where $F_B$ is the number of bicoloured cycles. We call $B$ \emph{planar} if $g(B)=0$. In rank three, melonic graphs are planar, but not every planar graph is melonic (Example~\ref{ex:Bstar}). For $p\geq 1$ let $\mathcal{M}_p$ be the disjoint union of $p$ melons; its invariant is $\Tr_{\mathcal{M}_p}=(T\cdot\bar T)^p$. We call $\mathcal{M}_2$ the \emph{double melon}.

\subsection{Gaussian measure, slots and Wick pairings}\label{ssec:gaussian}
Let $d\mu_N$ be the Gaussian probability measure on $\mathbb{C}^{N^D}$ with density proportional to $\exp(-N^{D-1}\,T\cdot\bar T)$, so that $\mathbb{E}[T_a\bar T_b]=N^{-(D-1)}\delta_{ab}$ and $\mathbb{E}[T_aT_b]=0$. This is the normalization of \cite{gurau2014universality,BTM}.

Given a finite list $X_1,\dots,X_q$ of $D$-coloured graphs, not necessarily connected, we write $r(X)$ for the number of connected components of $X$ and $p(X)$ for half its number of vertices. A \emph{slot} is a finite disjoint union of melonic graphs. We use slots from Section~\ref{sec:reduction} on. Given $B$ and $X_1,\dots,X_q$ with $m$ black and $m$ white vertices in total, labelled as one graph with permutations $(b_1,\dots,b_D)\in\mathfrak{S}_m^D$, a \emph{Wick pairing} is a bijection $\pi\in\mathfrak{S}_m$ contracting $j^\bullet$ with $\pi(j)^\circ$. We call these contractions \emph{propagators}, or edges of colour $0$. For each colour $c$, the permutation $\sigma_c=b_c^{-1}\pi$ maps black vertices to black vertices: it follows the propagator from $j^\bullet$ to $\pi(j)^\circ$ and then the colour-$c$ edge back to a black vertex.

\begin{figure}[ht]
\centering
\begin{subfigure}[b]{0.3\textwidth}
\centering
\begin{tikzpicture}
\coordinate (b1) at (0,1.6); \coordinate (w1) at (0,0);
\draw[col1] (b1) to[bend right=40] (w1);
\draw[col2] (b1) -- (w1);
\draw[col3] (b1) to[bend left=40] (w1);
\node[bv] at (b1) {}; \node[wv] at (w1) {};
\end{tikzpicture}
\caption{melon\\ $g=0,\ \psi_0=1,\ \ell=1$}
\end{subfigure}
\begin{subfigure}[b]{0.3\textwidth}
\centering
\begin{tikzpicture}[x=0.6cm]
\foreach \j in {1,2,3} { \coordinate (b\j) at (\j,1.6); \coordinate (w\j) at (\j,0); }
\foreach \j/\t in {1/1,2/2,3/3} \draw[col1] (b\j) -- (w\t);
\foreach \j/\t in {1/2,2/3,3/1} \draw[col2] (b\j) -- (w\t);
\foreach \j/\t in {1/3,2/1,3/2} \draw[col3] (b\j) -- (w\t);
\foreach \j in {1,2,3} { \node[bv] at (b\j) {}; \node[wv] at (w\j) {}; }
\end{tikzpicture}
\caption{$K_{3,3}$\\ $g=1,\ \psi_0=0,\ \ell=3$}
\end{subfigure}
\begin{subfigure}[b]{0.3\textwidth}
\centering
\begin{tikzpicture}[x=0.6cm]
\foreach \j in {1,2,3,4} { \coordinate (b\j) at (\j,1.6); \coordinate (w\j) at (\j,0); }
\foreach \j/\t in {1/1,2/2,3/3,4/4} \draw[col1] (b\j) -- (w\t);
\foreach \j/\t in {1/4,2/3,3/2,4/1} \draw[col2] (b\j) -- (w\t);
\foreach \j/\t in {1/3,2/4,3/1,4/2} \draw[col3] (b\j) -- (w\t);
\foreach \j in {1,2,3,4} { \node[bv] at (b\j) {}; \node[wv] at (w\j) {}; }
\end{tikzpicture}
\caption{$B_\star$ (planar, non-melonic)\\ $g=0,\ \psi_0=0,\ \ell=3$}
\end{subfigure}
\caption{Three invariants in rank three, drawn with black vertices on top and white vertices below, and the colour-$c$ edge from $j^\bullet$ to $b_c(j)^\circ$. In any melonic model their leading behaviour is $\ell(B)N^{\psi_0(B)}x^n$ by Corollary~\ref{cor:moments}. In (c) the bound $\psi_0\leq 1-g$ of \cite[(38) and Lemma~1]{gurau2014universality} is strict.}\label{fig:invariants}
\end{figure}
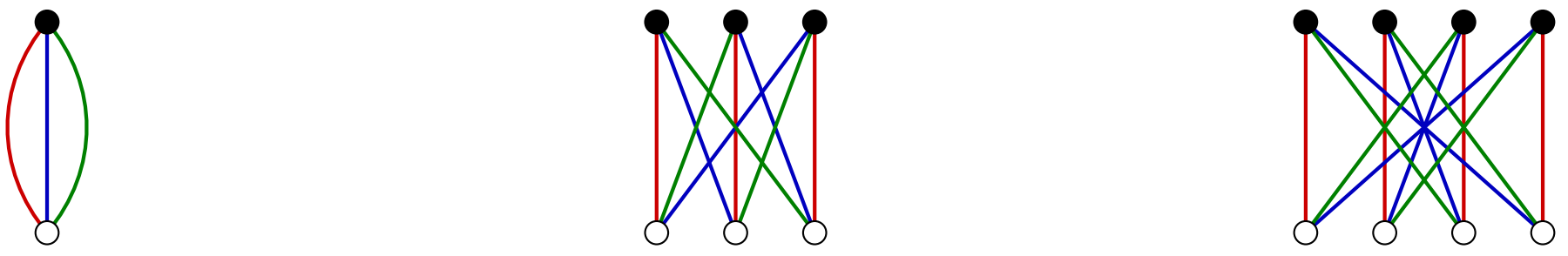

For a pairing $\pi\in\mathfrak{S}_m$ put $\sigma_c=b_c^{-1}\pi$. The cycles of $\sigma_c$ are the \emph{faces} of colour $(0,c)$, and $F(\pi)=\sum_{c=1}^D\mathrm{cyc}(\sigma_c)$ is their number (Fig.~\ref{fig:wick}). We call $B$ and the $X_s$ the \emph{blocks}. A pairing is \emph{block-connected} if the graph whose vertices are the blocks, and whose edges are the propagators, is connected (Fig.~\ref{fig:blocks}).

\par\medskip\noindent\begin{minipage}{\linewidth}\captionsetup{type=figure}
\centering
\begin{subfigure}[b]{0.3\textwidth}
\centering
\begin{tikzpicture}[x=0.75cm,y=0.75cm]
\foreach \j in {1,2,3} { \coordinate (b\j) at (\j,2.2); \coordinate (w\j) at (\j,0); }
\foreach \j/\t in {1/1,2/2,3/3} \draw[lbc1] (b\j) -- (w\t);
\foreach \j/\t in {1/2,2/3,3/1} \draw[lbc2] (b\j) -- (w\t);
\foreach \j/\t in {1/3,2/1,3/2} \draw[lbc3] (b\j) -- (w\t);
\foreach \j/\t in {1/2,2/1,3/3} \draw[lbprop] (b\j) to[bend right=24] (w\t);
\foreach \j in {1,2,3} {
  \node[lbbv, label={[lblab]above:$\j^\bullet$}] at (b\j) {};
  \node[lbwv, label={[lblab]below:$\j^\circ$}] at (w\j) {};}
\end{tikzpicture}
\caption{$K_{3,3}$ with $\pi=(1\,2)$}
\end{subfigure}%
\begin{subfigure}[b]{0.66\textwidth}
\centering
\begin{tikzpicture}[x=0.5cm,y=0.5cm]
\newcommand{\facepanel}[6]{%
  \begin{scope}[shift={(#1,#2)}]
    \foreach \j/\t in {#4} \draw[#3] (\j,1.9) -- (\t,0);
    \foreach \j/\t in {#5} \draw[lbprop] (\j,1.9) to[bend right=24] (\t,0);
    \foreach \j in {1,2,3} { \node[lbbv] at (\j,1.9) {}; \node[lbwv] at (\j,0) {}; }
    \node[lbelab] at (2,-0.75) {#6};
  \end{scope}}
\foreach \c/\x in {1/0,2/4.2,3/8.4} \node[lblab] at (\x+2,3.0) {colour $(0,\c)$};
\facepanel{0}{0}{lbc1}{1/1,2/2,3/3}{1/2,2/1,3/3}{$2$}
\facepanel{4.2}{0}{lbc2}{1/2,2/3,3/1}{1/2,2/1,3/3}{$2$}
\facepanel{8.4}{0}{lbc3}{1/3,2/1,3/2}{1/2,2/1,3/3}{$2$}
\facepanel{0}{-4.2}{lbc1}{1/1,2/2,3/3}{1/1,2/2,3/3}{$3$}
\facepanel{4.2}{-4.2}{lbc2}{1/2,2/3,3/1}{1/1,2/2,3/3}{$1$}
\facepanel{8.4}{-4.2}{lbc3}{1/3,2/1,3/2}{1/1,2/2,3/3}{$1$}
\node[lblab, anchor=east] at (0.3,0.95)  {$\pi=(1\,2)$};
\node[lblab, anchor=east] at (0.3,-3.25) {$\pi=\mathrm{id}$};
\node[lblab, anchor=west] at (11.9,0.95)  {$F=6$};
\node[lblab, anchor=west] at (11.9,-3.25) {$F=5$};
\end{tikzpicture}
\caption{faces of each colour, and their number}
\end{subfigure}
\captionsetup{font=footnotesize}
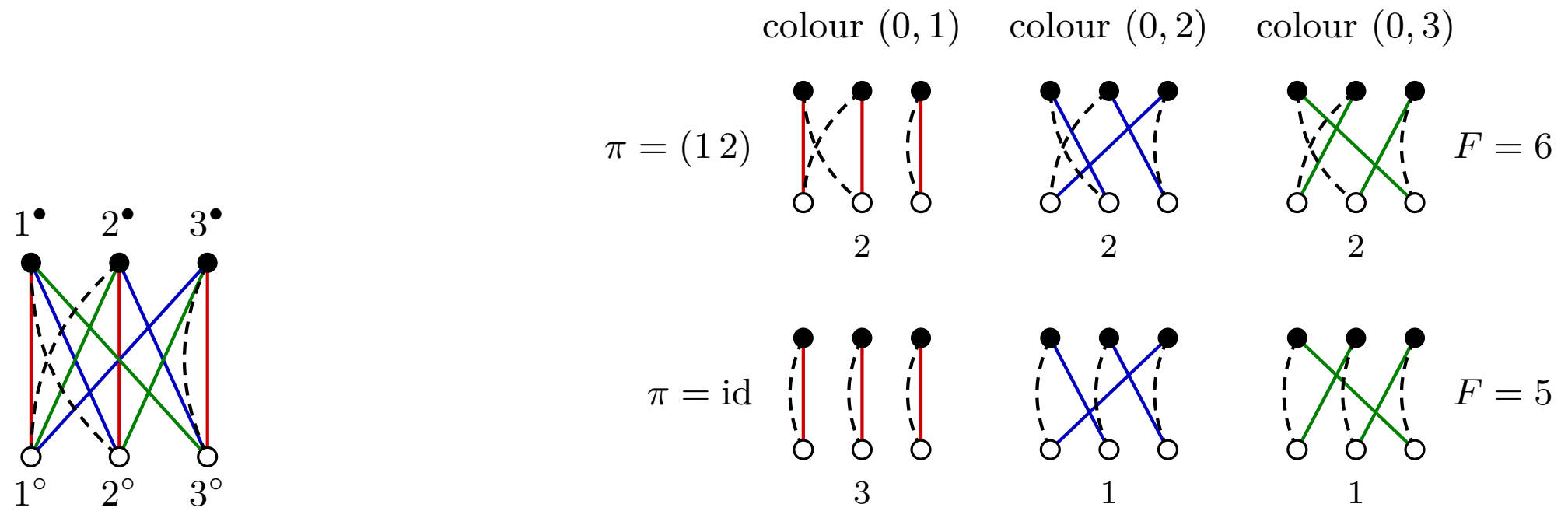
\caption{Wick pairings of $K_{3,3}$ in rank $3$; colour $1$ is red, $2$ blue and $3$ green, and the dashed propagators join $j^\bullet$ to $\pi(j)^\circ$. (a) The pairing $\pi=(1\,2)$. (b) For $\pi=(1\,2)$ and $\pi=\mathrm{id}$, the propagators together with the colour-$c$ edges, for each $c$. The number below each panel is its number of cycles, and $F$ is their total.}\label{fig:wick}
\end{minipage}\par\medskip

\begin{lemma}[Wick and diagram formulas]\label{lem:wick}\label{lem:diagram}
We have
\begin{equation*}
\mathbb{E}\Bigl[\Tr_B\prod_s\Tr_{X_s}\Bigr]=\sum_{\pi\in\mathfrak{S}_m}N^{F(\pi)-(D-1)m},\qquad \kappa\bigl(\Tr_B,\Tr_{X_1},\dots,\Tr_{X_q}\bigr)=\sum_{\pi\ \text{block-connected}}N^{F(\pi)-(D-1)m}.
\end{equation*}
\end{lemma}

\begin{proof}
The first formula is \cite[\S4.3, (37)]{gurau2014universality} with $\sigma=1$, applied to the disjoint union of $B$ and the $X_s$, whose invariant is $\Tr_B\prod_s\Tr_{X_s}$. Covering graphs there are our pairings $\pi$, with $\sum_iF^{0i}=F(\pi)$. It is Wick's theorem \cite[Ch.~1]{janson1997gaussian} with one free index per face. The second follows from the Leonov--Shiryaev formula \cite{leonov1959method}, \cite[Ch.~3 and Ch.~7]{peccati2011wiener}, since the only non-zero joint cumulants of the entries are $\kappa(T_a,\bar T_b)=N^{-(D-1)}\delta_{ab}$.
\end{proof}

\par\medskip\noindent\begin{minipage}{\linewidth}\captionsetup{type=figure}
\centering
\newcommand{\blockpanel}[2]{%
\begin{tikzpicture}[x=0.8cm,y=0.8cm,
  pr/.style={out=0,in=180}, pl/.style={out=180,in=0}, pb/.style={bend right=24}, ps/.style={}]
\coordinate (b0) at (0,1.6); \coordinate (w0) at (0,0);
\coordinate (k1) at (2.2,1.6); \coordinate (k2) at (3.4,1.6);
\coordinate (w1) at (2.2,0);   \coordinate (w2) at (3.4,0);
\coordinate (k3) at (5.6,1.6); \coordinate (w3) at (5.6,0);
\foreach \l/\a/\b in {B/-0.55/0.55, X_1/1.65/3.95, X_2/5.05/6.15} {
  \draw[lbbox] (\a,-0.45) rectangle (\b,2.05);
  \node[lblab, anchor=south] at ({(\a+\b)/2},2.05) {$\l$};}
\draw[lbc1] (b0) to[bend right=50] (w0); \draw[lbc2] (b0) -- (w0); \draw[lbc3] (b0) to[bend left=50] (w0);
\draw[lbc2] (k1) to[bend right=40] (w1); \draw[lbc3] (k1) to[bend left=40] (w1);
\draw[lbc2] (k2) to[bend right=40] (w2); \draw[lbc3] (k2) to[bend left=40] (w2);
\draw[lbc1] (k1) -- (w2); \draw[lbc1] (k2) -- (w1);
\draw[lbc1] (k3) to[bend right=50] (w3); \draw[lbc2] (k3) -- (w3); \draw[lbc3] (k3) to[bend left=50] (w3);
\foreach \s/\t/\o in {#1} \draw[lbprop] (\s) to[\o] (\t);
\foreach \v in {b0,k1,k2,k3} \node[lbbv] at (\v) {};
\foreach \v in {w0,w1,w2,w3} \node[lbwv] at (\v) {};
\foreach \n/\x in {B/0, X/2.8, Y/5.6} \coordinate (g\n) at (\x,-1.45);
\foreach \s/\t in {#2} \draw[line width=0.5pt] (g\s) -- (g\t);
\foreach \n/\l in {B/B, X/X_1, Y/X_2}
  \node[circle, draw=black, fill=white, line width=0.5pt, inner sep=0.5pt, minimum size=13pt, font=\scriptsize] at (g\n) {$\l$};
\end{tikzpicture}}
\begin{subfigure}[b]{0.48\textwidth}
\centering
\blockpanel{b0/w1/pr, k1/w0/pl, k2/w3/pr, k3/w2/pl}{B/X, X/Y}
\caption{block-connected}
\end{subfigure}\hfill
\begin{subfigure}[b]{0.48\textwidth}
\centering
\blockpanel{b0/w0/pb, k1/w1/ps, k2/w3/pr, k3/w2/pl}{X/Y}
\caption{not block-connected}
\end{subfigure}
\captionsetup{font=footnotesize}
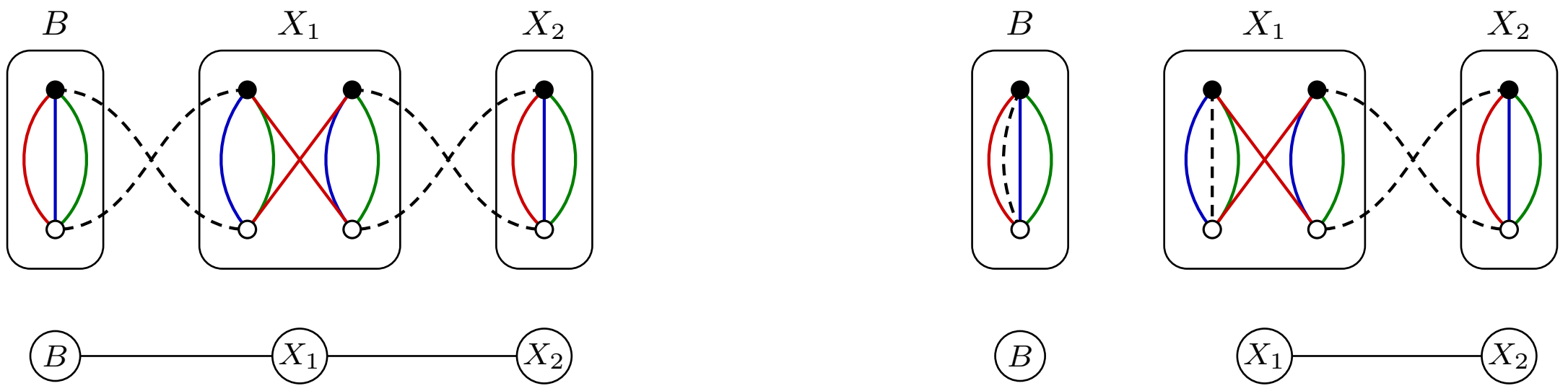
\caption{Two Wick pairings of the blocks $B=$ melon, $X_1=P_1$ and $X_2=$ melon in rank $3$, with black vertices on top and dashed propagators. Below each panel is its block graph: one vertex per block, and one edge per propagator joining two different blocks. (a) A block-connected pairing. (b) A pairing that is not block-connected.}\label{fig:blocks}
\end{minipage}\par\medskip

In particular $\int^{\mathrm{conn}}$ in \cite{BTM} is a joint cumulant. For example, take the blocks $B=$ melon, $X_1=P_1$ and $X_2=$ melon of Fig.~\ref{fig:blocks}. The pairing in panel (a) links all three blocks and therefore contributes to $\kappa(\Tr_B,\Tr_{X_1},\Tr_{X_2})$. The pairing in (b) pairs $B$ with itself, which leaves $B$ isolated in the block graph: it contributes to $\mathbb{E}[\Tr_B]\,\kappa(\Tr_{X_1},\Tr_{X_2})$ and not to the joint cumulant. For arbitrary $D$-coloured graphs $X_1,\dots,X_q$ we define
\begin{equation}\label{eq:IN}
\mathcal{I}_N[B;X_1,\dots,X_q]=N^{(D-1)q-1-\sum_j(r(X_j)-1)}\,\kappa\bigl(\Tr_B,\Tr_{X_1},\dots,\Tr_{X_q}\bigr)=\sum_{\pi\ \text{block-conn.}}N^{\psi(\pi)-1},
\end{equation}
with $m=n+\sum_jp(X_j)$ and
\begin{equation}\label{eq:psi}
\psi(\pi)=F(\pi)-(D-1)m+(D-1)q-\sum_j\bigl(r(X_j)-1\bigr).
\end{equation}
For connected $X_j$ this is the quantity of Theorem~\ref{thm:main}. For a graph with $r$ components, the factor $N^{-(r-1)}$ makes $\mathcal{M}_p$ stand for $(T\cdot\bar T)^p/N^{p-1}$. For $q=0$, $\mathcal{I}_N[B]=N^{-1}\mathbb{E}[\Tr_B]$ and
\begin{equation*}
\psi_0(B)=\max_{\pi\in\mathfrak{S}_n}\bigl(F(\pi)-(D-1)n\bigr),\qquad \ell(B)=\#\{\pi:F(\pi)-(D-1)n=\psi_0(B)\}.
\end{equation*}

\begin{remark}[Gur\u{a}u's convergence order]\label{rem:gurau}
Gur\u{a}u's Gaussian measure \cite[(8)]{gurau2014universality} with $\sigma=1$ is ours. His covering graphs of $B$ are $B$ together with the propagators of a Wick pairing. They are thus in bijection with Wick pairings, and the minimal covering graphs are those maximizing $F(\pi)$ \cite[Def.~11]{gurau2014universality}. Comparing \cite[(10) and (42)]{gurau2014universality} with the expansion $N^{-1}\mathbb{E}[\Tr_B]=\ell(B)N^{\psi_0(B)-1}+O(N^{\psi_0(B)-2})$, the convergence order of $B$ is $\Omega(B)=1-\psi_0(B)$ and the number $R(B)$ of minimal covering graphs is $\ell(B)$. Thus $\psi_0$ and $\ell$ are the constants of Gur\u{a}u's universality theorem, which are hard to determine in general~\cite[\S4.3.2]{gurau2014universality}.
\end{remark}

\begin{example}\label{ex:Bstar}
The planar non-melonic graph $B_\star$ with $b_1=\id$, $b_2=(1\,4)(2\,3)$, $b_3=(1\,3)(2\,4)$ (Fig.~\ref{fig:invariants}(c)) has $g(B_\star)=0$ but $\psi_0(B_\star)=0$ and $\ell(B_\star)=3$. The graph $K_{3,3}$ ($b_c(j)=j+c-1\bmod 3$) has $g=1$, $\psi_0=0$ and $\ell=3$. Figure~\ref{fig:wick} shows the count. For each colour $c$, the propagators and the colour-$c$ edges form the cycles of $\sigma_c$. The transposition $(1\,2)$ has two faces of each colour; hence $F=6=(D-1)n$ and it contributes $N^0$. The identity has $3+1+1=5$ faces and contributes $N^{-1}$. The three transpositions are the maximizing pairings.
\end{example}
\subsection{Arbitrary blocks}\label{ssec:blocks}
More generally, let $Y_0,\dots,Y_q$ be $D$-coloured graphs, not necessarily connected, with $m$ black vertices in total. We call them \emph{blocks}, define Wick pairings, faces and block-connectivity as in Section~\ref{ssec:gaussian}, and put
\begin{equation}\label{eq:chi}
\chi(\pi)=F(\pi)-(D-1)m-\sum_{i=0}^q\bigl(r(Y_i)-1\bigr).
\end{equation}
The second formula of Lemma~\ref{lem:diagram} does not need connected blocks and therefore gives
\begin{equation}\label{eq:diagchi}
N^{-\sum_i(r(Y_i)-1)}\,\kappa\bigl(\Tr_{Y_0},\dots,\Tr_{Y_q}\bigr)=\sum_{\pi\ \text{block-conn.}}N^{\chi(\pi)}.
\end{equation}
Let $\Psi(Y_0;\dots;Y_q)$ be the largest $\chi(\pi)$ over block-connected pairings and $\Lambda(Y_0;\dots;Y_q)$ the number of pairings attaining it, with $\Psi=-\infty$ and $\Lambda=0$ if there are none. For connected $Y_0=B$ and arbitrary $Y_j=X_j$, comparing \eqref{eq:psi} with \eqref{eq:chi} gives $\chi=\psi-(D-1)q$ and hence
\begin{equation}\label{eq:INchi}
\mathcal{I}_N[B;X_1,\dots,X_q]=N^{(D-1)q-1}\sum_{\pi\ \text{block-conn.}}N^{\chi(\pi)},
\end{equation}
a Laurent polynomial in $N$ with top exponent $\Psi(B;X_1,\dots,X_q)+(D-1)q-1$ and top coefficient $\Lambda(B;X_1,\dots,X_q)$. Since every pairing contributes a monomial with coefficient $+1$, no cancellation occurs. We use $\psi$ in statements and $\chi$ in proofs. Unlike \eqref{eq:psi}, \eqref{eq:chi} corrects every block by its number of components; all blocks thus play the same role.

\section{Melonic reduction}\label{sec:reduction}

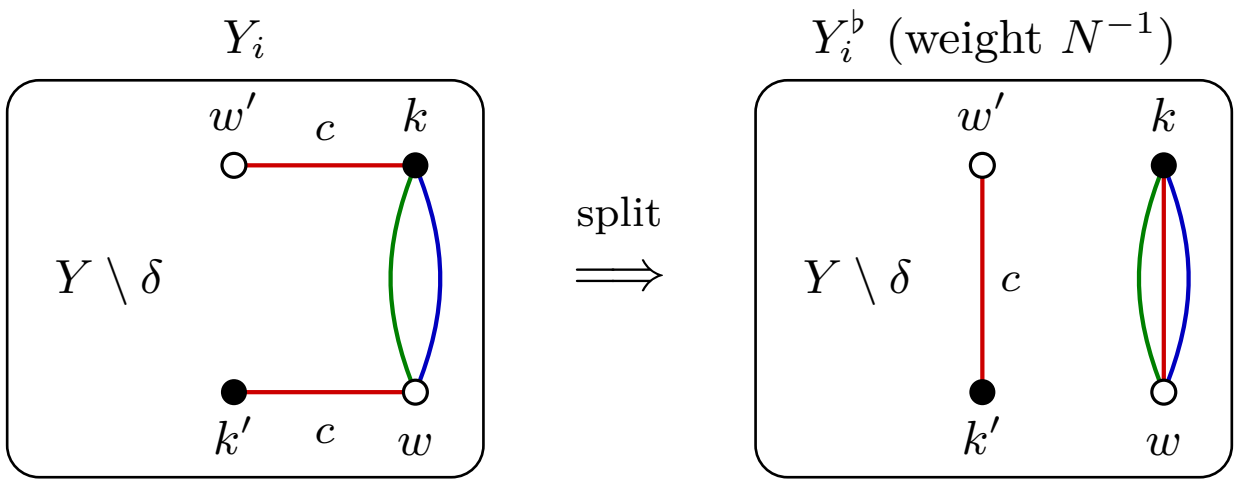
\begin{figure}[ht]
\centering
\begin{tikzpicture}[scale=0.75]
\foreach \X/\after in {0/0,6.6/1} {
\begin{scope}[xshift=\X cm]
\draw[lbbox] (-2.0,-0.75) rectangle (2.2,2.75);
\node[lblab] at (-1.1,1) {$Y\setminus\delta$};
\coordinate (wp) at (0,2); \coordinate (kp) at (0,0);
\coordinate (k) at (1.6,2); \coordinate (w) at (1.6,0);
\ifnum\after=0
  \node[lblab, anchor=south] at (0.1,2.8) {$Y_i$};
  \draw[lbc1] (wp) -- node[lbelab, above] {$c$} (k); \draw[lbc1] (kp) -- node[lbelab, below] {$c$} (w);
  \draw[lbc3] (k) to[bend right=22] (w); \draw[lbc2] (k) to[bend left=22] (w);
\else
  \node[lblab, anchor=south] at (0.1,2.8) {$Y_i^\flat$ (weight $N^{-1}$)};
  \draw[lbc1] (wp) -- node[lbelab, right] {$c$} (kp);
  \draw[lbc3] (k) to[bend right=22] (w); \draw[lbc1] (k) -- (w); \draw[lbc2] (k) to[bend left=22] (w);
\fi
\node[lbwv, label={[lblab]above:$w'$}] at (wp) {};
\node[lbbv, label={[lblab]below:$k'$}] at (kp) {};
\node[lbbv, label={[lblab]above:$k$}] at (k) {};
\node[lbwv, label={[lblab]below:$w$}] at (w) {};
\end{scope}}
\node[lbelab, anchor=south] at (3.4,1.25) {split};
\node[font=\normalsize] at (3.4,1) {$\Longrightarrow$};
\end{tikzpicture}
\captionsetup{font=footnotesize}
\caption{Dipole splitting, drawn in rank $3$ with $c$ the red colour. The $(D-1)$-dipole $(k,w)$ of the melonic component $Y$ becomes a separate melon, and the colour-$c$ edges are rewired from $kw'$, $k'w$ to $kw$, $k'w'$. Only colour $c$ changes, by a transposition, so the exponent is unchanged exactly when $k$ and $k'$ lie on the same colour-$(0,c)$ face (Lemma~\ref{lem:split}); the extra component costs a factor $N^{-1}$.}\label{fig:splitting}
\end{figure}

Dipoles also drive \cite[Lemma~3.1]{TwofoldUniversality}, where a maximal pairing contracts each dipole with a gain of $D-2$ faces. Here we split dipoles off melonic graphs instead, inside an arbitrary block of a cumulant. Let $Y_0,\dots,Y_q$ be blocks as in Section~\ref{ssec:blocks}, suppose that the block $Y_i$ contains a connected melonic component $Y$ with at least four vertices, and let $\delta=(k,w)$ be a $(D-1)$-dipole of $Y$ missing the colour $c$ such that $Y\setminus\delta$ is melonic; by (M1) such a dipole exists. Put $w'=b_c(k)$ and $k'=b_c^{-1}(w)$. Since $Y$ is connected with at least four vertices, $w'\neq w$ and $k'\neq k$. We compare three blocks (Fig.~\ref{fig:splitting}):
\begin{itemize}
\item $Y_i$, the original block;
\item $Y_i^\flat$, on the same vertices, in which $Y$ is replaced by $(Y\setminus\delta)\sqcup\{\text{melon on }(k,w)\}$, that is $b_c^\flat(k)=w$ and $b_c^\flat(k')=w'$;
\item $Y_i''$, on the vertices of $Y_i$ other than $k,w$, in which $Y$ is replaced by $Y\setminus\delta$.
\end{itemize}
The block $Y_i^\flat$ has one more component than $Y_i$, while $Y_i''$ has as many. If $Y$ is a component of a slot $Z\subseteq Y_i$, the corresponding parts $Z^\flat$ of $Y_i^\flat$ and $Z''$ of $Y_i''$ are again slots, since $Y\setminus\delta$ is melonic. We write $\chi$, $\chi^\flat$ and $\chi''$ for the exponents \eqref{eq:chi} computed with $Y_i$, $Y_i^\flat$ and $Y_i''$, the other blocks being unchanged.

\begin{lemma}[Splitting]\label{lem:split}
For every pairing $\pi$, block-connectivity is the same for $Y_i$ and $Y_i^\flat$. Moreover $\chi^\flat(\pi)=\chi(\pi)$ if $k$ and $k'$ lie in the same cycle of $\sigma_c=b_c^{-1}\pi$, and $\chi^\flat(\pi)=\chi(\pi)-2$ otherwise. We call $\pi$ \emph{good} in the first case and \emph{bad} in the second.
\end{lemma}

\begin{proof}
Block-connectivity is unchanged because the blocks have the same vertices. The permutations agree except $b_c^\flat=b_c\circ(k\,k')$, and hence $\sigma_c^\flat=(k\,k')\circ\sigma_c$. Multiplying by a transposition increases the number of cycles by one if $k,k'$ share a cycle and decreases it by one otherwise. The extra component of $Y_i^\flat$ contributes $-1$ to \eqref{eq:chi}.
\end{proof}

\begin{lemma}[Bad pairings]\label{lem:bad}
Let $\pi$ be block-connected and bad. Then $\pi(k)\neq w$. Put $y=\pi^{-1}(w)$ and $u=\pi(k)$, and define $\pi'$ on the vertices other than $k,w$ by $\pi'(j)=\pi(j)$ for $j\notin\{k,y\}$ and $\pi'(y)=u$. Then $\pi'$ is block-connected for the blocks with $Y_i''$ in place of $Y_i$, and $\chi(\pi)=\chi''(\pi')-(D-2)$.
\end{lemma}

\begin{proof}
If $\pi(k)=w$ then $\sigma_c(k)=b_c^{-1}(w)=k'$ and $\pi$ is good. So $\pi(k)\neq w$, hence $y\neq k$ and $u\neq w$ (Fig.~\ref{fig:lbreroute}).
\par\smallskip\noindent\begin{minipage}{\linewidth}\centering
\begin{tikzpicture}[scale=0.75]
\foreach \X/\after in {0/0,8.4/1} {
\begin{scope}[xshift=\X cm]
\draw[lbbox] (-0.6,-0.75) rectangle (2.2,2.75);
\draw[lbbox] (4.0,-0.75) rectangle (5.6,2.75);
\node[lblab, anchor=south] at (4.8,2.8) {other blocks};
\coordinate (wp) at (0,2); \coordinate (kp) at (0,0);
\coordinate (k) at (1.6,2); \coordinate (w) at (1.6,0);
\coordinate (u) at (4.8,2); \coordinate (y) at (4.8,0);
\draw[lbc1] (wp) -- node[lbelab, right] {$c$} (kp);
\ifnum\after=0
  \node[lblab, anchor=south] at (0.8,2.8) {$Y_i^\flat$};
  \draw[lbc3] (k) to[bend right=22] (w); \draw[lbc1] (k) -- (w); \draw[lbc2] (k) to[bend left=22] (w);
  \draw[lbprop, lbmid] (k) -- node[lbelab, above] {$\pi$} (u);
  \draw[lbprop, lbmid] (y) -- node[lbelab, below] {$\pi$} (w);
  \node[lbbv, label={[lblab]above:$k$}] at (k) {};
  \node[lbwv, label={[lblab]below:$w$}] at (w) {};
\else
  \node[lblab, anchor=south] at (0.8,2.8) {$Y_i''$};
  \draw[lbprop, lbmid] (y) -- node[lbelab, left] {$\pi'$} (u);
\fi
\node[lbwv, label={[lblab]above:$w'$}] at (wp) {};
\node[lbbv, label={[lblab]below:$k'$}] at (kp) {};
\node[lbwv, label={[lblab]above:$u$}] at (u) {};
\node[lbbv, label={[lblab]below:$y$}] at (y) {};
\end{scope}}
\node[font=\normalsize] at (6.7,1) {$\Longrightarrow$};
\end{tikzpicture}
\captionsetup{font=footnotesize}
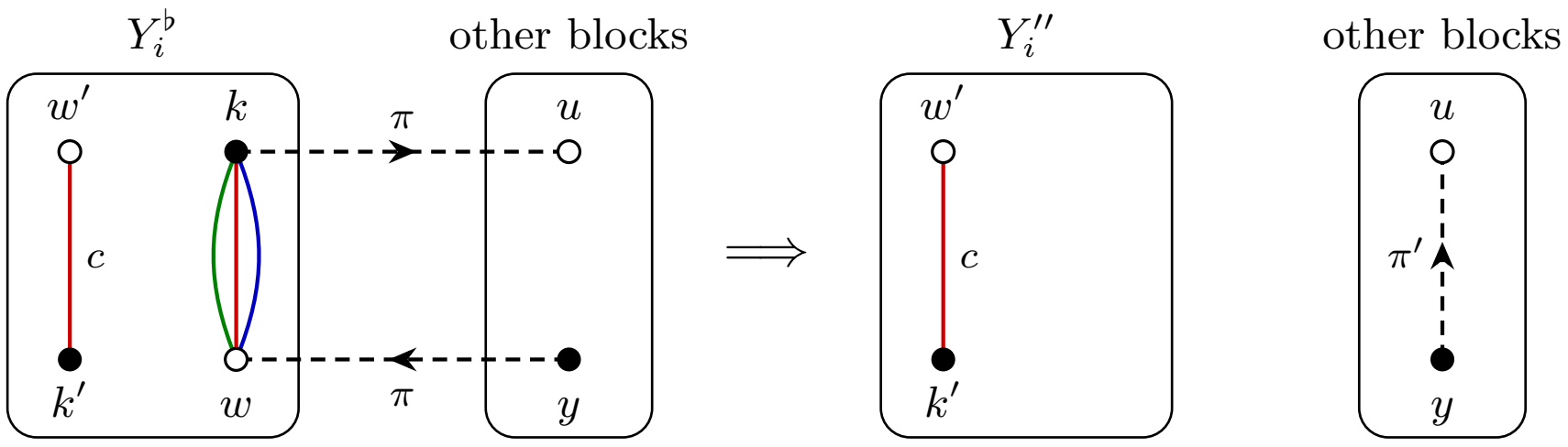
\captionof{figure}{The rerouting in Lemma~\ref{lem:bad}, drawn in rank $3$ with $c$ the red colour. Left: a bad pairing $\pi$ on $Y_i^\flat$, where $(k,w)$ is a melon and the colour-$c$ edge joins $k'$ to $w'$. Right: $\pi'$ on $Y_i''$, where the melon is deleted and the propagators $y\to w$ and $k\to u$ are replaced by $y\to u$. Dashed lines are propagators.}\label{fig:lbreroute}
\end{minipage}\par\smallskip

\emph{Faces.} We compare $\pi$ on $Y_i^\flat$ with $\pi'$ on $Y_i''$. In $Y_i^\flat$ the pair $(k,w)$ is a melon, that is, $b_e^\flat(k)=w$ for every $e\in[D]$, and deleting it leaves exactly $Y_i''$. Hence $(b_e'')^{-1}=(b_e^\flat)^{-1}$ on every white vertex other than $w$. For every colour $e$ we have $\sigma_e^\flat(y)=(b_e^\flat)^{-1}(w)=k$ and $\sigma_e^\flat(k)=(b_e^\flat)^{-1}(u)$, while $\sigma_e'(y)=(b_e'')^{-1}(u)=\sigma_e^\flat(k)$. For $j\notin\{k,y\}$ we have $\pi'(j)=\pi(j)\notin\{w,u\}$ and therefore $\sigma_e'(j)=\sigma_e^\flat(j)$. Thus $\sigma_e'$ is $\sigma_e^\flat$ with $k$ removed from its cycle (Fig.~\ref{fig:lbface}). Since that cycle contains $y\neq k$, no cycle disappears and $F''(\pi')=F^\flat(\pi)$. Passing from $(Y_i^\flat,\pi)$ to $(Y_i'',\pi')$ removes one propagator and one component of $Y_i$ and leaves the other blocks unchanged. Then \eqref{eq:chi} gives $\chi''(\pi')=\chi^\flat(\pi)+(D-1)+1=\chi^\flat(\pi)+D$. By Lemma~\ref{lem:split}, $\chi^\flat(\pi)=\chi(\pi)-2$, since $\pi$ is bad. Hence $\chi(\pi)=\chi''(\pi')-(D-2)$.
\par\smallskip\noindent\begin{minipage}{\linewidth}\centering
\begin{tikzpicture}[scale=0.75]
\node[lblab, anchor=east] at (-0.5,0) {$\pi$ on $Y_i^\flat$};
\coordinate (y) at (0,0); \coordinate (w) at (1.6,0); \coordinate (k) at (3.2,0);
\coordinate (u) at (4.8,0); \coordinate (v) at (6.4,0); \coordinate (e) at (8.0,0);
\draw[lbprop, lbmid] (y) -- node[lbelab, above] {$\pi$} (w);
\draw[lbc2, lbmid] (w) -- node[lbelab, above] {$e$} (k);
\draw[lbprop, lbmid] (k) -- node[lbelab, above] {$\pi$} (u);
\draw[lbc2, lbmid] (u) -- node[lbelab, above] {$e$} (v);
\draw[lbwalk, lbmid] (v) -- (e);
\foreach \P/\L in {y/y,k/k,v/\sigma_e(k)} \node[lbbv, label={[lblab]below:$\L$}] at (\P) {};
\foreach \P/\L in {w/w,u/u} \node[lbwv, label={[lblab]below:$\L$}] at (\P) {};
\node[lblab, anchor=east] at (-0.5,-1.6) {$\pi'$ on $Y_i''$};
\coordinate (y2) at (0,-1.6); \coordinate (u2) at (4.8,-1.6); \coordinate (v2) at (6.4,-1.6); \coordinate (e2) at (8.0,-1.6);
\draw[lbprop, lbmid] (y2) -- node[lbelab, above] {$\pi'$} (u2);
\draw[lbc2, lbmid] (u2) -- node[lbelab, above] {$e$} (v2);
\draw[lbwalk, lbmid] (v2) -- (e2);
\foreach \P/\L in {y2/y,v2/\sigma_e(k)} \node[lbbv, label={[lblab]below:$\L$}] at (\P) {};
\node[lbwv, label={[lblab]below:$u$}] at (u2) {};
\end{tikzpicture}
\captionsetup{font=footnotesize}
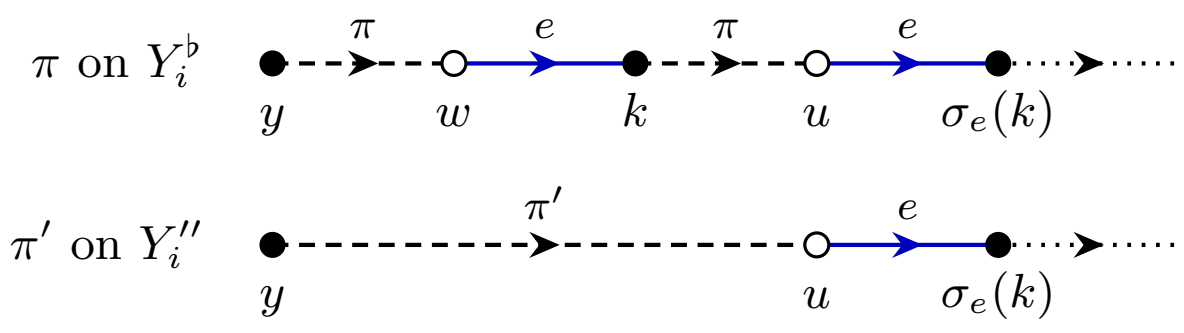
\captionof{figure}{A face of colour $e$, for any $e\in[D]$ (drawn in blue), followed along $\sigma_e$ in the direction of the arrows; dotted lines are further steps. Under $\pi'$ the face loses $k$ and is otherwise unchanged, so $F''(\pi')=F^\flat(\pi)$.}\label{fig:lbface}
\end{minipage}\par\smallskip

\emph{Connectivity.} Write $\mathrm{bl}(j)$ for the block containing $j$. Passing to $\pi'$ replaces the propagators $y\to w$ and $k\to u$ by $y\to u$; here $k,w$ lie in $Y_i$. If $\mathrm{bl}(y)=Y_i$ or $\mathrm{bl}(u)=Y_i$, the new propagator joins the same two blocks as one of the removed ones, and the other removed one joins $Y_i$ to itself, so the block graph keeps the same connections. Otherwise suppose that the block graph of $\pi'$ is disconnected, and let $A$ be the set of blocks in the component containing $Y_i$. Since $y\to u$ is a propagator of $\pi'$, the blocks $\mathrm{bl}(y)$ and $\mathrm{bl}(u)$ are both in $A$ or both outside $A$. If both were in $A$, no propagator of $\pi$ would join $A$ to the remaining blocks, contradicting block-connectivity of $\pi$. Hence $\mathrm{bl}(y),\mathrm{bl}(u)\notin A$. The only propagators of $\pi$ joining $A$ to the other blocks are therefore $y\to w$ and $k\to u$. We follow the orbit of $\sigma_c=b_c^{-1}\pi$ from $k$ (Fig.~\ref{fig:lborbit}). Since $\sigma_c(j)$ lies in the block of $\pi(j)$, the orbit starts at $\sigma_c(k)=b_c^{-1}(u)\notin A$, and it can re-enter $A$ only along a propagator from outside $A$ into $A$, that is through $y\to w$, after which the next point is $\sigma_c(y)=b_c^{-1}(w)=k'$. As the orbit returns to $k\in A$, it passes through $k'$, and then $\pi$ would be good, a contradiction.
\end{proof}
\par\smallskip\noindent\begin{minipage}{\linewidth}\centering
\begin{tikzpicture}[scale=0.75]
\draw[lbbox] (-0.6,-0.75) rectangle (3.8,2.75);
\draw[lbbox] (5.8,-0.75) rectangle (10.2,2.75);
\node[lblab, anchor=south] at (1.6,2.8) {blocks in $A$};
\node[lblab, anchor=south] at (8.0,2.8) {blocks not in $A$};
\coordinate (wp) at (0,2);  \coordinate (k) at (3.2,2);
\coordinate (kp) at (0,0);  \coordinate (w) at (3.2,0);
\coordinate (u) at (6.4,2); \coordinate (v) at (9.6,2); \coordinate (y) at (6.4,0); \coordinate (c) at (9.6,0);
\draw[lbc3] (k) to[bend right=22] (w); \draw[lbc2] (k) to[bend left=22] (w);
\draw[lbprop, lbmid] (k) -- node[lbelab, above] {$\pi$} (u);
\draw[lbc1, lbmid] (u) -- node[lbelab, above] {$c$} (v);
\draw[lbwalk, lbmid] (v) -- (c); \draw[lbwalk, lbmid] (c) -- (y);
\draw[lbprop, lbmid] (y) -- node[lbelab, below] {$\pi$} (w);
\draw[lbc1, lbmid] (w) -- node[lbelab, below] {$c$} (kp);
\draw[lbwalk, lbmid] (kp) -- (wp);
\draw[lbc1, lbmid] (wp) -- node[lbelab, above] {$c$} (k);
\node[lbbv, label={[lblab]above:$k$}] at (k) {};
\node[lbwv, label={[lblab]above:$w'$}] at (wp) {};
\node[lbwv, label={[lblab]above:$u$}] at (u) {};
\node[lbbv, label={[lblab]above:$\sigma_c(k)$}] at (v) {};
\node[lbwv, label={[lblab]below:$w$}] at (w) {};
\node[lbbv, label={[lblab]below:$k'$}] at (kp) {};
\node[lbbv, label={[lblab]below:$y$}] at (y) {};
\end{tikzpicture}
\captionsetup{font=footnotesize}
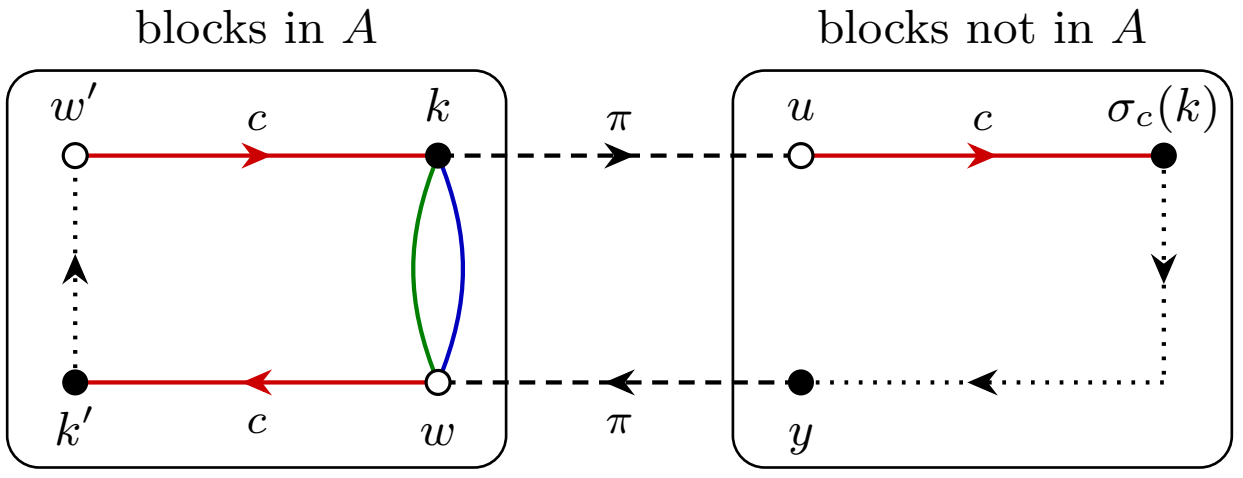
\captionof{figure}{The orbit of $k$ under $\sigma_c$ for the original block $Y_i$, where the colour-$c$ edges (red) join $k$ to $w'$ and $w$ to $k'$, and the other colours of the dipole $(k,w)$ are blue and green. If $\pi'$ were not block-connected, the orbit could leave $A$ only along $k\to u$ and return only along $y\to w$, after which it reaches $k'$; then $\pi$ would be good.}\label{fig:lborbit}
\end{minipage}\par\smallskip

\begin{proposition}[Melonic reduction]\label{prop:melred}
Let $Y_0,\dots,Y_q$ be blocks, and suppose that $Y_i=Z\sqcup R$, where $Z$ is a slot with $p(Z)=p$ and $R$ is arbitrary, possibly empty. Replacing $Z$ by $\mathcal{M}_p$ changes neither $\Psi$ nor $\Lambda$, whatever the other blocks are.
\end{proposition}

\begin{proof}
We write $\Psi_p$ and $\Lambda_p$ for the values with $\mathcal{M}_p\sqcup R$ in place of $Y_i$, the other blocks being fixed.

\emph{Monotonicity.} We show $\Psi_{p-1}\leq\Psi_p$ for $p\geq 2$. Given a block-connected pairing for $\mathcal{M}_{p-1}\sqcup R$, add one melon to the block and pair its two vertices with each other. The pairing stays block-connected, and since $F$ grows by $D$, $m$ by $1$ and $r(Y_i)$ by $1$, $\chi$ is unchanged.

\emph{Induction.} We induct on the number of vertices of $Z$ lying in components with at least four vertices. If there are none, $Z=\mathcal{M}_p$. Otherwise we choose a component $Y$ of $Z$ with at least four vertices and a dipole $\delta$ of $Y$ by (M1). Then $Z^\flat$ and $Z''$ are slots with $p(Z^\flat)=p$ and $p(Z'')=p-1\geq1$, both with fewer vertices in components with at least four vertices. By induction, $\Psi^\flat=\Psi_p$, $\Lambda^\flat=\Lambda_p$ and $\Psi''=\Psi_{p-1}$. If $\Psi_p=-\infty$, then $\Psi_{p-1}=-\infty$ by monotonicity, so by Lemma~\ref{lem:bad} there are no bad block-connected pairings and the inequalities below are vacuous. Let $\pi$ be block-connected. If $\pi$ is good, then $\chi(\pi)=\chi^\flat(\pi)\leq\Psi_p$. If $\pi$ is bad, then Lemma~\ref{lem:bad} and monotonicity give
\begin{equation*}
\chi(\pi)=\chi''(\pi')-(D-2)\leq\Psi_{p-1}-(D-2)<\Psi_p.
\end{equation*}
In this case Lemma~\ref{lem:split} also gives $\chi^\flat(\pi)=\chi(\pi)-2<\Psi_p$. Hence the pairings with $\chi(\pi)=\Psi_p$ are exactly the good pairings with $\chi^\flat(\pi)=\Psi_p$, which are all the pairings with $\chi^\flat(\pi)=\Psi_p$. This gives $\Psi=\Psi_p$ and $\Lambda=\Lambda_p$.
\end{proof}

Proposition~\ref{prop:melred} acts on any block, including the observable, and it needs no a priori bound on the exponents. Applied once to each interaction, with $R$ empty, it gives the following.

\begin{corollary}[Reduction principle]\label{cor:reduction}
Let $B$ be a $D$-coloured graph and $X_1,\dots,X_q$ slots with $p(X_j)=p_j$. Then $\Psi$ and $\Lambda$ take the same values for $(B;X_1,\dots,X_q)$ and for $(B;\mathcal{M}_{p_1},\dots,\mathcal{M}_{p_q})$.
\end{corollary}

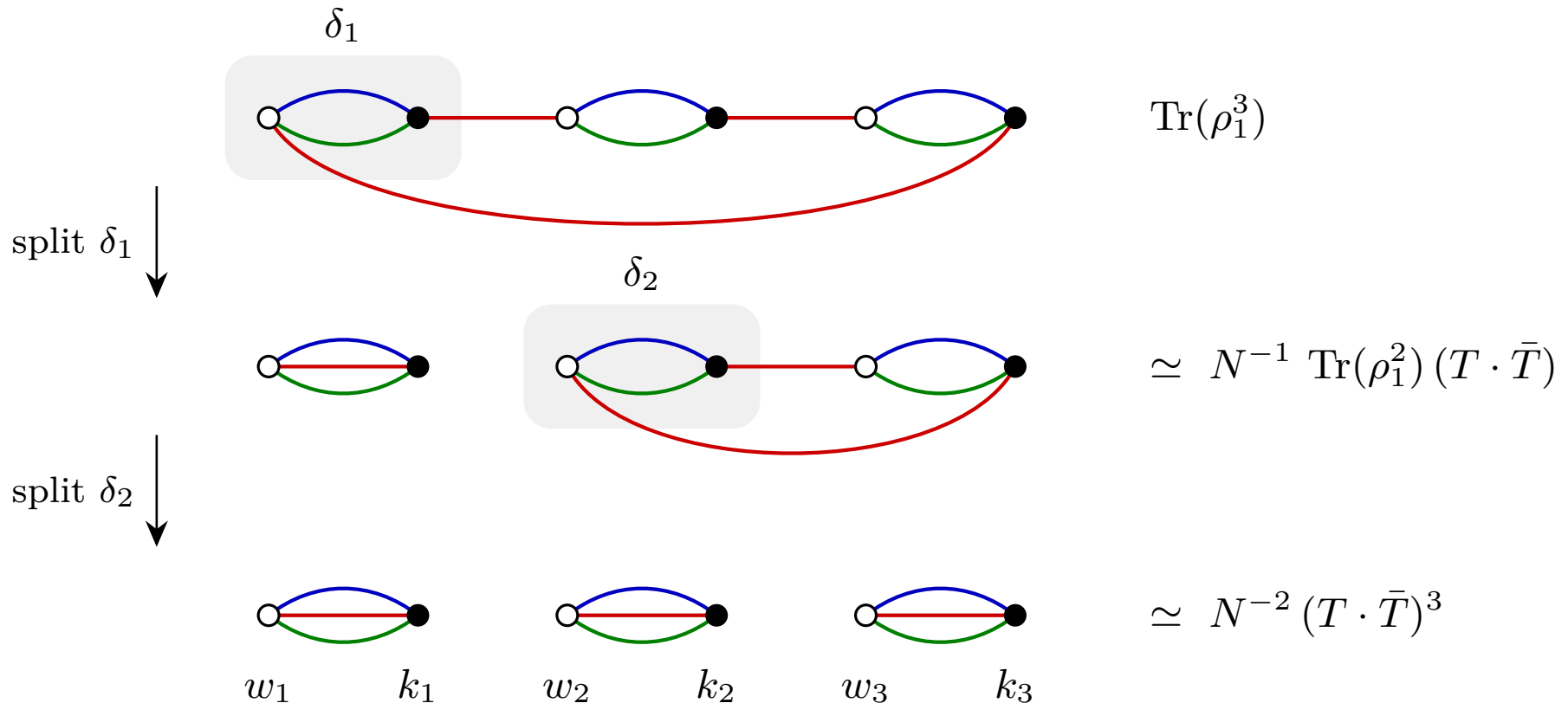
\begin{figure}[t]
\centering
\begin{tikzpicture}[x=0.95cm,y=0.95cm]
\def\rowgap{2.0}
\newcommand{\dipole}[3]{%
  \draw[lbc2] (#1,#2) to[bend left=38]  (#1+1.2,#2);
  \draw[lbc3] (#1,#2) to[bend right=38] (#1+1.2,#2);
  \ifnum#3=1 \draw[lbc1] (#1,#2) -- (#1+1.2,#2); \fi}
\newcommand{\verts}[1]{%
  \foreach \x in {0,2.4,4.8} {
    \node[lbwv] at (\x,#1) {};
    \node[lbbv] at (\x+1.2,#1) {};}}
\fill[gray!12, rounded corners=6pt] (-0.35,-0.5) rectangle (1.55,0.5);
\fill[gray!12, rounded corners=6pt] (2.05,-\rowgap-0.5) rectangle (3.95,-\rowgap+0.5);
\dipole{0}{0}{0} \dipole{2.4}{0}{0} \dipole{4.8}{0}{0}
\draw[lbc1] (1.2,0) -- (2.4,0);
\draw[lbc1] (3.6,0) -- (4.8,0);
\draw[lbc1] (0,0) to[out=-62, in=-118, looseness=0.55] (6.0,0);
\dipole{0}{-\rowgap}{1} \dipole{2.4}{-\rowgap}{0} \dipole{4.8}{-\rowgap}{0}
\draw[lbc1] (3.6,-\rowgap) -- (4.8,-\rowgap);
\draw[lbc1] (2.4,-\rowgap) to[out=-62, in=-118, looseness=0.75] (6.0,-\rowgap);
\dipole{0}{-2*\rowgap}{1} \dipole{2.4}{-2*\rowgap}{1} \dipole{4.8}{-2*\rowgap}{1}
\verts{0} \verts{-\rowgap} \verts{-2*\rowgap}
\node[lblab, anchor=south] at (0.6,0.5)          {$\delta_1$};
\node[lblab, anchor=south] at (3.0,-\rowgap+0.5) {$\delta_2$};
\foreach \i/\x in {1/0,2/2.4,3/4.8} {
  \node[lblab, anchor=north] at (\x,-2*\rowgap-0.32)     {$w_{\i}$};
  \node[lblab, anchor=north] at (\x+1.2,-2*\rowgap-0.32) {$k_{\i}$};}
\foreach \r/\d in {0/1,1/2} {
  \draw[-{Stealth[length=2mm, width=1.7mm]}, line width=0.5pt]
    (-0.9,-\r*\rowgap-0.55) -- (-0.9,-\r*\rowgap-\rowgap+0.55);
  \node[lbelab, anchor=east] at (-0.95,-\r*\rowgap-0.5*\rowgap) {split $\delta_{\d}$};}
\node[lblab, anchor=west] at (7.0,0)          {$\Tr(\rho_1^3)$};
\node[lblab, anchor=west] at (7.0,-\rowgap)   {$\simeq\ N^{-1}\,\Tr(\rho_1^2)\,(T\cdot\bar T)$};
\node[lblab, anchor=west] at (7.0,-2*\rowgap) {$\simeq\ N^{-2}\,(T\cdot\bar T)^3$};
\end{tikzpicture}
\captionsetup{font=footnotesize}
\caption{Melonic reduction of the chain $\Tr(\rho_1^3)$ in rank $3$; colour $1$ is red, $2$ blue and $3$ green. The dipole $\delta_i=(k_i,w_i)$ shaded in a row is the one split off to give the next row. The right-hand column gives the invariant of each row.}\label{fig:cascade}
\end{figure}

Figure~\ref{fig:cascade} follows the induction in the proof of Proposition~\ref{prop:melred} for the chain $\Tr(\rho_1^3)$, whose dipoles $\delta_i=(k_i,w_i)$ are joined by the colours $\neq1$. Each step splits one dipole off as in Fig.~\ref{fig:splitting}. Only colour-$1$ edges are rewired, the vertices do not move, and the new component costs a factor $N^{-1}$ in \eqref{eq:chi}. After two steps the slot is $\mathcal{M}_3$. We write $\simeq$ when two graphs give the same $\Psi$ and $\Lambda$ in every cumulant in which they are a block. Then $\Tr(\rho_1^3)\simeq N^{-1}\Tr(\rho_1^2)\,(T\cdot\bar T)\simeq N^{-2}(T\cdot\bar T)^3$: at each step good pairings keep their exponent (Lemma~\ref{lem:split}) and bad ones drop by $D-2$ (Lemma~\ref{lem:bad}).

\section{The radial variable}\label{sec:radial}

This section treats the radius $M=T\cdot\bar T$. For interactions that are disjoint unions of melons, the cumulant factorizes exactly into Gaussian data of $B$ and a cumulant of a single Gamma variable (Section~\ref{ssec:radial}), and a loop equation for that variable computes the limit (Section~\ref{ssec:loop}). The same loop equation, with the direction of $T$ integrated out, governs the two-point function of an arbitrary model (Section~\ref{ssec:interacting}). Nothing here uses Section~\ref{sec:reduction}.

\subsection{Radial reduction}\label{ssec:radial}
We set $M=T\cdot\bar T$. Under $d\mu_N$ the $|T_a|^2$ are i.i.d.\ exponential with mean $N^{-(D-1)}$, hence $M\sim\mathrm{Gamma}(N^D,\ \text{rate }N^{D-1})$. The direction $\Theta=T/\sqrt{M}$ is uniform on the unit sphere and independent of $M$. Since $\Tr_B$ has bidegree $(n,n)$, we have $\Tr_B=M^nZ_B$ with $Z_B=\Tr_B(\Theta,\bar\Theta)$ independent of $M$. In particular $\Tr_{\mathcal{M}_p}=M^p$.

\begin{lemma}\label{lem:radial}
For all $p_1,\dots,p_q\geq 1$,
\begin{equation*}
\kappa\bigl(\Tr_B,M^{p_1},\dots,M^{p_q}\bigr)=\mathbb{E}[Z_B]\,\kappa\bigl(M^n,M^{p_1},\dots,M^{p_q}\bigr),\qquad \mathbb{E}[\Tr_B]=\mathbb{E}[Z_B]\,\mathbb{E}[M^n].
\end{equation*}
\end{lemma}

\begin{proof}
We expand the joint cumulant by the moment--cumulant formula,
\begin{equation*}
\kappa(Y_0,\dots,Y_q)=\sum_{\mathcal{P}}(-1)^{|\mathcal{P}|-1}(|\mathcal{P}|-1)!\prod_{V\in\mathcal{P}}\mathbb{E}\Bigl[\prod_{j\in V}Y_j\Bigr],
\end{equation*}
with $Y_0=M^nZ_B$ and $Y_j=M^{p_j}$. By independence, the block $V$ containing $0$ contributes
\begin{equation*}
\mathbb{E}[Z_B]\,\mathbb{E}\Bigl[M^n\prod_{j\in V\setminus\{0\}}M^{p_j}\Bigr],
\end{equation*}
and no other block involves $Z_B$.
\end{proof}

With $\mathcal{K}=N^D$ and $\xi=M/N$ we have $\xi\sim\mathrm{Gamma}(\mathcal{K},\ \text{rate }\mathcal{K})$. The slots $\mathcal{M}_{p_j}$ have $r(\mathcal{M}_{p_j})-1=p_j-1$ in \eqref{eq:IN}, and Lemma~\ref{lem:radial} gives
\begin{align*}
\mathcal{I}_N\bigl[B;\mathcal{M}_{p_1},\dots,\mathcal{M}_{p_q}\bigr]&=N^{(D-1)q-1-\sum_j(p_j-1)}\,\mathbb{E}[Z_B]\,\kappa\bigl(M^n,M^{p_1},\dots,M^{p_q}\bigr)\\
&=N^{Dq-1-\sum_jp_j}\,\frac{\mathbb{E}[\Tr_B]}{N^n\,\mathbb{E}[\xi^n]}\,N^{n+\sum_jp_j}\,\kappa\bigl(\xi^n,\xi^{p_1},\dots,\xi^{p_q}\bigr),
\end{align*}
that is, the exact identity
\begin{equation}\label{eq:radial}
\mathcal{I}_N\bigl[B;\mathcal{M}_{p_1},\dots,\mathcal{M}_{p_q}\bigr]=N^{-1}\mathbb{E}[\Tr_B]\cdot\frac{\mathcal{K}^q\,\kappa(\xi^n;\xi^{p_1},\dots,\xi^{p_q})}{\mathbb{E}[\xi^n]}.
\end{equation}

\subsection{A loop equation for Gamma variables}\label{ssec:loop}
Let $\mathcal{K}>0$ and $\xi\sim\mathrm{Gamma}(\mathcal{K},\ \text{rate }\mathcal{K})$. Then $\mathbb{E}[\xi^r]=\prod_{i=0}^{r-1}(1+i/\mathcal{K})$ and therefore
\begin{equation}\label{eq:gamma}
\mathcal{K}\,\mathbb{E}[\xi^{r+1}]=\mathcal{K}(1+r/\mathcal{K})\prod_{i=0}^{r-1}(1+i/\mathcal{K}) = (\mathcal{K}+r)\,\mathbb{E}[\xi^r],\qquad r\geq 0.
\end{equation}
Fix integers $p_1,\dots,p_q\geq 1$ and formal variables $t=(t_1,\dots,t_q)$. For $\beta\in\mathbb{N}^q$ write $\beta\cdot p=\sum_j\beta_jp_j$, and write $(\xi^p)^{(\beta)}$ for the list in which $\xi^{p_j}$ appears $\beta_j$ times, as before Corollary~\ref{cor:moments}. For $\nu\geq 0$ define the formal power series
\begin{equation*}
A_\nu(t)=\sum_{\beta\in\mathbb{N}^q}\frac{(-\mathcal{K})^{|\beta|}\,t^\beta}{\beta!}\,\mathbb{E}\bigl[\xi^{\nu+\beta\cdot p}\bigr],\qquad a_\nu=\frac{A_\nu}{A_0}.
\end{equation*}
Formally $A_\nu=\mathbb{E}\bigl[\xi^\nu e^{-\mathcal{K}\sum_jt_j\xi^{p_j}}\bigr]$. The coefficients of $A_\nu$ are polynomials in $\mathcal{K}$ and $1/\mathcal{K}$, and since $A_0$ has constant term $1$, $a_\nu$ is well defined. The joint cumulants of random variables $Y,Y_1,\dots,Y_q$ with finite moments are the coefficients of the formal series $\log\mathbb{E}\bigl[\exp(uY+\sum_js_jY_j)\bigr]$. Hence the coefficient of $s^\beta/\beta!$ in
\begin{equation*}
\frac{\mathbb{E}\bigl[Y\exp(\sum_js_jY_j)\bigr]}{\mathbb{E}\bigl[\exp(\sum_js_jY_j)\bigr]}=\partial_u\log\mathbb{E}\Bigl[\exp\Bigl(uY+\sum_js_jY_j\Bigr)\Bigr]\Big|_{u=0}
\end{equation*}
is the joint cumulant of $Y$ and the list in which $Y_j$ appears $\beta_j$ times. Taking $Y=\xi^\nu$, $Y_j=\xi^{p_j}$ and $s=-\mathcal{K}t$ gives, by definition,
\begin{equation}\label{eq:coef}
[t^\beta]\,a_\nu=\frac{(-\mathcal{K})^{|\beta|}}{\beta!}\,\kappa\bigl(\xi^\nu;(\xi^p)^{(\beta)}\bigr).
\end{equation}
Both sides vanish for $\nu=0$ and $\beta\neq 0$, since $a_0=1$.

\begin{lemma}[Loop equation]\label{lem:loop}
For every $\nu\geq 0$,
\begin{equation*}
\nu\,a_\nu=\mathcal{K}\Bigl(a_{\nu+1}-a_\nu+\sum_{j=1}^qp_jt_j\,a_{\nu+p_j}\Bigr).
\end{equation*}
\end{lemma}

\begin{proof}
We apply \eqref{eq:gamma} with $r=\nu+\beta\cdot p$, multiply by $(-\mathcal{K})^{|\beta|}t^\beta/\beta!$ and sum over $\beta$. The factor $\beta\cdot p$ produces $\sum_jp_jt_j\partial_{t_j}A_\nu$. This gives
\begin{equation*}
\nu A_\nu+\sum_jp_jt_j\partial_{t_j}A_\nu=\mathcal{K}(A_{\nu+1}-A_\nu).
\end{equation*}
Shifting $\beta_j$ by one shows $t_j\partial_{t_j}A_\nu=-\mathcal{K}t_jA_{\nu+p_j}$. Substituting and dividing by $A_0$ gives the claim.
\end{proof}

No integration by parts or analytic property of the Laplace transform is needed: the loop equation is \eqref{eq:gamma}, read coefficientwise.

\begin{lemma}\label{lem:poly}
For all $\nu\geq 0$ and $\beta\in\mathbb{N}^q$, the coefficient $[t^\beta]\,a_\nu$ is a polynomial in $1/\mathcal{K}$. Consequently, for all integers $d_0\geq 0$ and $d_1,\dots,d_q\geq 1$, $\mathcal{K}^q\kappa(\xi^{d_0},\xi^{d_1},\dots,\xi^{d_q})$ is a polynomial in $1/\mathcal{K}$. In particular, $\kappa(\xi^{d_0},\dots,\xi^{d_q})=O(\mathcal{K}^{-q})$.
\end{lemma}

\begin{proof}
Lemma~\ref{lem:loop} gives $a_{\nu+1}=(1+\nu/\mathcal{K})\,a_\nu-\sum_jp_jt_j\,a_{\nu+p_j}$. Taking coefficients,
\begin{equation*}
[t^\beta]\,a_{\nu+1}=\Bigl(1+\frac{\nu}{\mathcal{K}}\Bigr)[t^\beta]\,a_\nu-\sum_{j:\,\beta_j\geq 1}p_j\,[t^{\beta-e_j}]\,a_{\nu+p_j},
\end{equation*}
where $e_j$ is the $j$-th unit vector. Induction on $|\beta|$ and, for fixed $\beta$, on $\nu$, starting from $a_0=1$, proves the first claim. The second follows by taking $p_j=d_j$, $\nu=d_0$ and $\beta=(1,\dots,1)$ in \eqref{eq:coef}.
\end{proof}

\begin{lemma}[Lagrange inversion]\label{lem:lagrange}
Let $p_1,\dots,p_L\geq 1$, and let $x(g)$ be the formal power series solution of $x+\sum_ip_ig_ix^{p_i}=1$, $x(0)=1$. For every $n\geq 1$ and $\alpha\in\mathbb{N}^L$,
\begin{equation*}
[g^\alpha]\,x(g)^n=\frac{(-1)^{|\alpha|}}{\alpha!}\,\mathrm{Mel}_{|\alpha|+1}\bigl(n,p^{(\alpha)}\bigr).
\end{equation*}
\end{lemma}

\begin{proof}
Both sides equal $1$ for $\alpha=0$. We write $x=1+y$ and $g_i=u\theta_i$. Then $y=u\varphi(y)$ with $\varphi(\lambda)=-\sum_ip_i\theta_i(1+\lambda)^{p_i}$, a power series over the field $\mathbb{Q}(\theta_1,\dots,\theta_L)$ with $\varphi(0)\neq 0$. By Lagrange--B\"urmann inversion \cite[\S5.4]{stanley1999enumerative}, for $q\geq 1$,
\begin{equation*}
[u^q]\,(1+y)^n=\frac1q\,[\lambda^{q-1}]\,n(1+\lambda)^{n-1}\varphi(\lambda)^q.
\end{equation*}
Let $|\alpha|=q$ and let $S=\sum_i\alpha_ip_i$ be the sum of the entries of $p^{(\alpha)}$. The coefficient of $\theta^\alpha$ in $\varphi(\lambda)^q$ is $(-1)^q\frac{q!}{\alpha!}\prod_ip_i^{\alpha_i}(1+\lambda)^S$. Since $g^\alpha=u^q\theta^\alpha$,
\begin{equation*}
[g^\alpha]\,x^n=\frac{(-1)^q}{\alpha!}\,(q-1)!\,n\prod_ip_i^{\alpha_i}\binom{n+S-1}{q-1}=\frac{(-1)^q}{\alpha!}\,n\prod_ip_i^{\alpha_i}\,\frac{(n+S-1)!}{(n+S-q)!},
\end{equation*}
which is $(-1)^q\mathrm{Mel}_{q+1}(n,p^{(\alpha)})/\alpha!$.
\end{proof}

\begin{proposition}\label{prop:gammalimit}
For all integers $n,p_1,\dots,p_q\geq 1$,
\begin{equation*}
\lim_{\mathcal{K}\to\infty}\mathcal{K}^q\,\kappa\bigl(\xi^n;\xi^{p_1},\dots,\xi^{p_q}\bigr)=\mathrm{Mel}_{q+1}(n,p_1,\dots,p_q).
\end{equation*}
\end{proposition}

\begin{proof}
By Lemma~\ref{lem:poly} every coefficient of $a_\nu$ is a polynomial in $1/\mathcal{K}$; let $\bar a_\nu$ be the formal series of their constant terms. Dividing Lemma~\ref{lem:loop} by $\mathcal{K}$ and setting $1/\mathcal{K}=0$ coefficientwise gives
\begin{equation*}
\bar a_{\nu+1}=\bar a_\nu-\sum_jp_jt_j\,\bar a_{\nu+p_j},\qquad \bar a_0=1.
\end{equation*}
As in the proof of Lemma~\ref{lem:poly}, these relations determine all $\bar a_\nu$. They are satisfied by $x(t)^\nu$, where $x(t)$ is the series solution of $x+\sum_jp_jt_jx^{p_j}=1$, because $x^{\nu+1}-x^\nu+\sum_jp_jt_jx^{\nu+p_j}=x^\nu\bigl(x-1+\sum_jp_jt_jx^{p_j}\bigr)=0$. Hence $\bar a_\nu=x(t)^\nu$. Then \eqref{eq:coef} with $\beta=(1,\dots,1)$ gives
\begin{equation*}
\lim_{\mathcal{K}\to\infty}\mathcal{K}^q\,\kappa\bigl(\xi^n;\xi^{p_1},\dots,\xi^{p_q}\bigr)=(-1)^q\,[t_1\cdots t_q]\,x(t)^n=\mathrm{Mel}_{q+1}(n,p_1,\dots,p_q)
\end{equation*}
by Lemma~\ref{lem:lagrange} with $L=q$ and $\alpha=(1,\dots,1)$.
\end{proof}

For instance $\mathcal{K}^2\kappa(\xi;\xi^2,\xi^2)=16+40/\mathcal{K}+24/\mathcal{K}^2$, with limit $\mathrm{Mel}_3(1,2,2)=16$.

\begin{remark}[Saddle point]\label{rem:saddle}
Heuristically, the series $x(g)$ is the saddle point of the radial variable. If each interaction $X_i$ is replaced by $\mathcal{M}_{p_i}$ with the normalization $N^{-(p_i-1)}$ of \eqref{eq:IN}, then $V_g=\mathcal{K}\sum_ig_i\xi^{p_i}$ and the law of $\xi$ becomes proportional to $\xi^{-1}\exp\bigl(-\mathcal{K}[\xi-\log\xi+\sum_ig_i\xi^{p_i}]\bigr)$. Its critical point as $\mathcal{K}\to\infty$ solves $1-1/\xi+\sum_ip_ig_i\xi^{p_i-1}=0$, which is \eqref{eq:twopoint}. At $g=0$ the exponent $\xi-\log\xi$ has its minimum at $\xi=1$, where the factor $e^{-\mathcal{K}(\xi-\log\xi-1)}$ of the Gamma density concentrates as $\mathcal{K}$ grows. The interaction tilts the exponent and moves its minimum to $x(g)$. The proof above does not use this.
\end{remark}

\subsection{The radial variable in arbitrary models}\label{ssec:interacting}
Throughout this subsection and Section~\ref{ssec:arbitrary}, $X_1,\dots,X_L$, $n_i$, $s_i$ and $V_g$ are as in \eqref{eq:Vs}. We write $\mathbf{n}=(n_1,\dots,n_L)$, $s=(s_1,\dots,s_L)$, $X^{(\alpha)}$ for the list in which $X_i$ appears $\alpha_i$ times (as before Corollary~\ref{cor:moments}), and $M=T\cdot\bar T$, $\mathcal{K}=N^D$, $\xi=M/N$, $\Theta=T/\sqrt M$ as in Section~\ref{ssec:radial}. As in the proof of Corollary~\ref{cor:moments}, for every invariant $\mathcal{O}$,
\begin{equation}\label{eq:coefs}
[g^\alpha]\,\mathbb{E}_g[\mathcal{O}]=\frac{(-1)^{|\alpha|}}{\alpha!}\,N^{\alpha\cdot s}\,\kappa\bigl(\mathcal{O},\Tr_{X^{(\alpha)}}\bigr),
\end{equation}
where $\mathcal{O}$ enters the cumulant as a single variable.

As in Section~\ref{ssec:radial}, we write $\Tr_{X_i}=M^{n_i}Z_i$ with $Z_i=\Tr_{X_i}(\Theta,\bar\Theta)$. For $0\neq\beta\in\mathbb{N}^L$ put $d_\beta=\beta\cdot\mathbf{n}$ and
\begin{equation}\label{eq:tau}
\tau_\beta=\frac{(-1)^{|\beta|+1}}{\beta!\,\mathcal{K}}\,N^{\beta\cdot(s+\mathbf{n})}\,\kappa_\Theta\bigl(Z^{(\beta)}\bigr),
\end{equation}
where $\kappa_\Theta(Z^{(\beta)})$ is the joint cumulant of the list $Z^{(\beta)}$ under the uniform measure on the unit sphere. Since $V_g=\sum_ig_iN^{s_i+n_i}\xi^{n_i}Z_i$ and $\Theta$ is independent of $\xi$, expanding $\log\mathbb{E}_\Theta[e^{-V_g}]$ in joint cumulants of the $Z_i$ gives, as formal series in $g$ whose coefficients are polynomials in $\xi$,
\begin{equation}\label{eq:W}
\mathbb{E}_\Theta\bigl[e^{-V_g}\bigr]=e^{-W(\xi)},\qquad W(\xi)=\mathcal{K}\sum_{\beta\neq0}\tau_\beta\,g^\beta\,\xi^{d_\beta}.
\end{equation}
We set
\begin{equation*}
b_\nu=\mathbb{E}_g[\xi^\nu]=\frac{\mathbb{E}\bigl[\xi^\nu e^{-W(\xi)}\bigr]}{\mathbb{E}\bigl[e^{-W(\xi)}\bigr]},\qquad \xi\sim\mathrm{Gamma}(\mathcal{K},\text{rate }\mathcal{K}).
\end{equation*}
Fix $\alpha$ and work modulo the monomials $g^\gamma$ with $\gamma\not\leq\alpha$. Then only the finitely many $\beta$ with $0\neq\beta\leq\alpha$ enter, and $b_\nu$ is the series $a_\nu$ of Section~\ref{ssec:loop} with the degrees $d_\beta$ in place of the $p_j$, after the substitution $t_\beta=\tau_\beta g^\beta$, which is well defined since each $g^\beta$ has positive degree. Lemma~\ref{lem:loop} therefore gives, modulo these monomials,
\begin{equation}\label{eq:recs}
b_{\nu+1}=\Bigl(1+\frac{\nu}{\mathcal{K}}\Bigr)b_\nu-\sum_{0\neq\beta\leq\alpha}d_\beta\,\tau_\beta\,g^\beta\,b_{\nu+d_\beta},\qquad b_0=1.
\end{equation}

\section{Proofs and further results}\label{sec:proofs}

\subsection{Arbitrary observables in melonic models}
\begin{proof}[Proof of Theorem~\ref{thm:main}]
Both $\mathcal{I}_N[B;X_1,\dots,X_q]$ and $\mathcal{I}_N[B;\mathcal{M}_{p_1},\dots,\mathcal{M}_{p_q}]$ are sums of monomials in $N$ with coefficient $+1$, so by \eqref{eq:INchi} and Corollary~\ref{cor:reduction} they have the same top exponent and the same top coefficient. In \eqref{eq:radial}, $N^{-1}\mathbb{E}[\Tr_B]=\ell(B)N^{\psi_0(B)-1}+O(N^{\psi_0(B)-2})$. The second factor is a ratio of polynomials in $\mathcal{K}^{-1}=N^{-D}$ whose denominator $\mathbb{E}[\xi^n]=\prod_{i<n}(1+i/\mathcal{K})$ has constant term $1$; by Lemma~\ref{lem:poly} and Proposition~\ref{prop:gammalimit} it equals $\mathrm{Mel}_{q+1}(n,p_1,\dots,p_q)+O(N^{-D})$. Hence $\mathcal{I}_N[B;\mathcal{M}_{p_1},\dots,\mathcal{M}_{p_q}]$ has top exponent $\psi_0(B)-1$ and top coefficient $\ell(B)\mathrm{Mel}_{q+1}(n,p_1,\dots,p_q)>0$, as does $\mathcal{I}_N[B;X_1,\dots,X_q]$. Since its exponents are integers, the remainder is $O(N^{\psi_0(B)-2})$.
\end{proof}

\begin{proof}[Proof of Corollary~\ref{cor:melonic}]
Apply Theorem~\ref{thm:main} with $B=C_0$, which has $n=p_0$ and $\psi_0=\ell=1$ by (M2).
\end{proof}

\begin{proof}[Proof of Corollary~\ref{cor:moments}]
Since $\mathbb{E}_g[\mathcal{O}]=\partial_u\log\mathbb{E}\bigl[e^{u\mathcal{O}-V_g}\bigr]\big|_{u=0}$, expanding the logarithm of this formal moment generating series in joint cumulants and keeping the terms linear in $u$ gives, as formal power series in $g$,
\begin{equation*}
\mathbb{E}_g[\Tr_B]=\sum_{q\geq 0}\frac{(-1)^q}{q!}\,\kappa\bigl(\Tr_B,\underbrace{V_g,\dots,V_g}_{q}\bigr).
\end{equation*}
By multilinearity of cumulants and \eqref{eq:IN},
\begin{equation*}
\mathbb{E}_g[\Tr_B]=N\sum_{q\geq 0}\frac{(-1)^q}{q!}\sum_{i_1,\dots,i_q=1}^Lg_{i_1}\cdots g_{i_q}\,\mathcal{I}_N[B;X_{i_1},\dots,X_{i_q}].
\end{equation*}
Fix $\alpha$ and put $q=|\alpha|$. The monomial $g^\alpha$ comes from the $q!/\alpha!$ tuples $(i_1,\dots,i_q)$ in which each $i$ occurs $\alpha_i$ times. For each of them $(p_{i_1},\dots,p_{i_q})$ is a rearrangement of $p^{(\alpha)}$. Since $\mathrm{Mel}_{q+1}$ is symmetric, Theorem~\ref{thm:main} gives
\begin{equation*}
[g^\alpha]\,\mathbb{E}_g[\Tr_B]=\frac{(-1)^q}{\alpha!}\,\ell(B)\,\mathrm{Mel}_{q+1}\bigl(n,p^{(\alpha)}\bigr)N^{\psi_0(B)}+O\bigl(N^{\psi_0(B)-1}\bigr).
\end{equation*}
By Lemma~\ref{lem:lagrange}, $(-1)^q\mathrm{Mel}_{q+1}(n,p^{(\alpha)})/\alpha!=[g^\alpha]\,x(g)^n$, which is non-zero.
\end{proof}

In the pillow model of Conjecture~\ref{conj:I}, $L=3$, $X_c=P_c$ and the couplings are $g/2$; the coefficient of $g^q$ in $\mathbb{E}_g[\Tr_B]$ is therefore $\frac{(-1)^q}{2^qq!}N\sum_{i_1,\dots,i_q}I_{i_1,\dots,i_q;N}[B]$, as stated in the introduction.

\begin{remark}[Subleading terms]\label{rem:subleading}
The subleading coefficients depend on the interactions. For instance, for $D=3$ and $B=\ominus$ the melon,
\begin{equation*}
I_{1,1;N}[\ominus]=16+\frac{40}{N}+\frac{16}{N^2}+\frac{8}{N^4},\qquad I_{1,2;N}[\ominus]=16+\frac{32}{N}+\frac{24}{N^2}+\frac{8}{N^3},
\end{equation*}
in agreement with \cite[\S5]{BTM}. In the proof, the interaction-sensitive contributions are the bad pairings, which are pushed down by $N^{-(D-2)}$.
\end{remark}

\begin{remark}[The rank]\label{rem:rank}
After Conjecture~I, the authors of \cite{BTM} remark that the number of tensor indices being three is essential. In the present proof the rank enters only through the factor $N^{-(D-2)}$ by which Lemma~\ref{lem:bad} suppresses bad pairings, so the conclusion holds for every $D\geq 3$, and the dividing line is $D=2$. There it fails. Taking $f=\Tr_X$ in the integration by parts of the proof of Proposition~\ref{prop:ward} (both identities below hold for every $D$) gives $\kappa(M,\Tr_X)=pN^{-(D-1)}\mathbb{E}[\Tr_X]$ for connected $X$ with $2p$ vertices, hence $\mathcal{I}_N[\ominus;X]=pN^{-1}\mathbb{E}[\Tr_X]$. For $D=2$ both pairings of $P_1$ have three faces, so $\mathcal{I}_N[\ominus;P_1]=2N^{-1}\cdot2N=4$, while \eqref{eq:radial} gives $\mathcal{I}_N[\ominus;\mathcal{M}_2]=\mathcal{K}\kappa(\xi;\xi^2)=2+2N^{-2}$ with $\mathcal{K}=N^2$. This is consistent with the non-Gaussian large-$N$ limits of matrix models \cite[\S3]{gurau2014universality}.
\end{remark}

\begin{remark}[Non-melonic interactions]\label{rem:nonmelonic}
The hypothesis that the interactions are melonic cannot be dropped. For $D=3$, the graph $K_{3,3}$ of Example~\ref{ex:Bstar} has six vertices. By Theorem~\ref{thm:main}, every melonic interaction $X$ with six vertices gives $\mathcal{I}_N[\ominus;X]=\mathrm{Mel}_2(1,3)+O(N^{-1})=3+O(N^{-1})$, whereas $\mathcal{I}_N[\ominus;K_{3,3}]$ is of order $N^{-1}$. Theorem~\ref{thm:main} still gives its leading term, with the roles exchanged. If $q=1$ and the graphs are connected, \eqref{eq:IN} is symmetric, $\mathcal{I}_N[\ominus;K_{3,3}]=\mathcal{I}_N[K_{3,3};\ominus]$. Theorem~\ref{thm:main} with observable $K_{3,3}$ ($n=3$, $\psi_0=0$, $\ell=3$) and the melon as interaction then gives $\ell(K_{3,3})\mathrm{Mel}_2(3,1)N^{-1}+O(N^{-2})=9N^{-1}+O(N^{-2})$. A single non-melonic member is thus controlled at its own scale.
\end{remark}

\subsection{Factorization of melonic observables}\label{ssec:arbitrary}
\begin{lemma}[The hypothesis]\label{rem:bounded}
For every $\alpha$, $[g^\alpha]\,N^{-1}\mathbb{E}_g[M]$ is $N^{\alpha\cdot s}$ times a Laurent polynomial in $N$ that does not depend on $s$, and it is bounded as $N\to\infty$ if and only if it converges. In particular, the hypothesis of Theorem~\ref{thm:arbitrary} implies that these coefficients converge.
\end{lemma}

\begin{proof}
The first claim follows from \eqref{eq:coefs} and Lemma~\ref{lem:diagram} (Lemma~\ref{lem:wickO} for real tensors). A finite sum $\sum_ec_eN^e$ with distinct real exponents is bounded as $N\to\infty$ if and only if $c_e=0$ for all $e>0$, in which case it converges to $c_0$.
\end{proof}

The hypothesis thus says that no scaling is too large. If all coefficients of positive order tend to $0$, then $m_2=1$ and Theorem~\ref{thm:arbitrary} reduces to the Gaussian case; the theorem has content at the largest admissible scalings.

\begin{theorem}\label{thm:factor}
Fix $\alpha$ and assume that $[g^{\alpha'}]\,N^{-1}\mathbb{E}_g[M]$ is bounded for every $\alpha'\leq\alpha$. Let $m_2$ be the series of the limits, defined modulo the monomials $g^\gamma$ with $\gamma\not\leq\alpha$. Then the following hold.
\begin{enumerate}[(i)]
\item For $0\neq\beta\leq\alpha$ the limits $\bar\tau_\beta=\lim_{N\to\infty}\tau_\beta$ exist.
\item For every $\nu\geq0$ and $\alpha'\leq\alpha$, $\lim_{N\to\infty}[g^{\alpha'}]\,\mathbb{E}_g[\xi^\nu]=[g^{\alpha'}]\,m_2^\nu$.
\item For all connected melonic $C_1,\dots,C_r$ with $2p_1,\dots,2p_r$ vertices and every $\alpha'\leq\alpha$, $\lim_{N\to\infty}[g^{\alpha'}]\,N^{-r}\mathbb{E}_g\bigl[\prod_{j=1}^r\Tr_{C_j}\bigr]=[g^{\alpha'}]\,m_2^{p_1+\dots+p_r}$.
\end{enumerate}
\end{theorem}

\begin{proof}
\emph{Convergence.} For $\alpha'=0$, $[g^0]b_\nu=\mathbb{E}[\xi^\nu]=\prod_{i<\nu}(1+i/\mathcal{K})\to1$. Let $0\neq\alpha'\leq\alpha$, and assume that $\tau_\beta$ converges for $0\neq\beta\leq\alpha'$, $\beta\neq\alpha'$, and that $[g^\beta]b_\nu$ converges for all such $\beta$, for $\beta=0$, and for all $\nu$. Taking $[g^{\alpha'}]$ in \eqref{eq:recs} at $\nu=0$ and isolating the term $\beta=\alpha'$,
\begin{equation*}
d_{\alpha'}\,\tau_{\alpha'}\,\mathbb{E}\bigl[\xi^{d_{\alpha'}}\bigr]=-[g^{\alpha'}]b_1-\sum_{\substack{0\neq\beta\leq\alpha'\ \beta\neq\alpha'}}d_\beta\,\tau_\beta\,[g^{\alpha'-\beta}]b_{d_\beta}.
\end{equation*}
Here $[g^{\alpha'}]b_1=[g^{\alpha'}]\,N^{-1}\mathbb{E}_g[M]$ converges by hypothesis and Lemma~\ref{rem:bounded}, the other terms converge by induction, and $\mathbb{E}[\xi^{d_{\alpha'}}]\to1$. Hence $\tau_{\alpha'}$ converges. Then \eqref{eq:recs}, with $[g^{\alpha'}]b_0=0$, shows by induction on $\nu$ that every $[g^{\alpha'}]b_\nu$ converges.

\emph{Limit.} We write $\bar b_\nu$ for the series of limits. They satisfy
\begin{equation*}
\bar b_{\nu+1}=\bar b_\nu-\sum_{\beta\neq0}d_\beta\,\bar\tau_\beta\,g^\beta\,\bar b_{\nu+d_\beta},\qquad \bar b_0=1,
\end{equation*}
which determine them by induction on the degree in $g$ and then on $\nu$, since every $\beta$ in the sum is non-zero. Consider, modulo the monomials $g^\gamma$ with $\gamma\not\leq\alpha$, the equation
\begin{equation}\label{eq:m2}
h+\sum_{\beta\neq0}d_\beta\,\bar\tau_\beta\,g^\beta\,h^{d_\beta}=1
\end{equation}
for a formal power series $h$. The same induction shows that \eqref{eq:m2} has a unique solution with constant term $1$. If $h$ is this solution, then $h^\nu$ satisfies the relations, because
\begin{equation*}
h^{\nu+1}-h^\nu+\sum_{\beta\neq0}d_\beta\,\bar\tau_\beta\,g^\beta\,h^{\nu+d_\beta}=h^\nu\Bigl(h-1+\sum_{\beta\neq0}d_\beta\,\bar\tau_\beta\,g^\beta\,h^{d_\beta}\Bigr)=0.
\end{equation*}
Hence $\bar b_\nu=h^\nu$ and $m_2=\bar b_1=h$. This proves (i) and (ii).

\emph{Melonic observables.} Fix $\alpha'\leq\alpha$ and put $S=p_1+\dots+p_r$. We take $\prod_j\Tr_{C_j}=\Tr_{Y_0}$, with $Y_0=C_1\sqcup\dots\sqcup C_r$, as one variable in \eqref{eq:coefs}. Since the interactions are connected, \eqref{eq:coefs} and \eqref{eq:diagchi} give
\begin{equation*}
[g^{\alpha'}]\,\mathbb{E}_g[\Tr_{Y_0}]=\frac{(-1)^{|\alpha'|}}{\alpha'!}\,N^{\alpha'\cdot s+r-1}\sum_{\pi}N^{\chi(\pi)},\qquad [g^{\alpha'}]\,\mathbb{E}_g\bigl[N^{r-S}M^S\bigr]=\frac{(-1)^{|\alpha'|}}{\alpha'!}\,N^{\alpha'\cdot s+r-1}\sum_{\pi}N^{\chi(\pi)},
\end{equation*}
the sums running over the block-connected pairings of the blocks $Y_0,X^{(\alpha')}$ and $\mathcal{M}_S,X^{(\alpha')}$ respectively. In the second, the factor $N^{r(\mathcal{M}_S)-1}=N^{S-1}$ of \eqref{eq:diagchi} combines with $N^{r-S}$ to give $N^{r-1}$. Applying Proposition~\ref{prop:melred} $r$ times, with $Z=C_j$ and $R$ the rest of the block, shows that the two sums have the same top exponent $\Psi$ and the same top coefficient $\Lambda$, and all their exponents are integers. If $\Lambda=0$ both coefficients vanish. Otherwise their difference is $O(N^{\alpha'\cdot s+r+\Psi-2})$, while the second coefficient has leading term $(-1)^{|\alpha'|}\Lambda N^{\alpha'\cdot s+r+\Psi-1}/\alpha'!$. The second coefficient equals $N^r[g^{\alpha'}]b_S$, since $N^{r-S}M^S=N^r\xi^S$; after division by $N^r$ it therefore converges by (ii). Hence $\alpha'\cdot s+\Psi\leq1$ and the difference divided by $N^r$ tends to $0$.
\end{proof}

\begin{remark}[The equation for $m_2$]\label{rem:m2}
The proof of Theorem~\ref{thm:factor} shows that $m_2$ is, modulo the monomials $g^\gamma$ with $\gamma\not\leq\alpha$, the unique solution of \eqref{eq:m2} with constant term $1$, that is, the unique series with $m_2(0)=1$ and
\begin{equation*}
m_2+\sum_{\beta\neq0}d_\beta\,\bar\tau_\beta\,g^\beta\,m_2^{d_\beta}=1.
\end{equation*}
Conversely, \eqref{eq:m2} determines the $\bar\tau_\beta$, $0\neq\beta\leq\alpha$, from $m_2$, recursively in $\beta$, since every $d_\beta$ is positive. Thus \eqref{eq:m2} reparametrizes $m_2$ by the limits $\bar\tau_\beta$ and does not constrain it. We use it in the converse direction, to read off the $\bar\tau_\beta$ once $m_2$ is known (Corollary~\ref{cor:overlap} and Proposition~\ref{prop:tetra}).
\end{remark}

\begin{proof}[Proof of Theorem~\ref{thm:arbitrary}]
The coefficients of $N^{-1}\mathbb{E}_g[M]$ converge by Lemma~\ref{rem:bounded}, and $m_2(0)=1$ since $N^{-1}\mathbb{E}[M]=\mathbb{E}[\xi]=1$. Since $\alpha$ is arbitrary, Theorem~\ref{thm:factor}(iii) gives the claim.
\end{proof}

\begin{proposition}[Ward identity]\label{prop:ward}
As formal power series in $g$, exactly,
\begin{equation*}
N^{D-1}\,\mathbb{E}_g[M]+\sum_{i=1}^Ln_ig_iN^{s_i}\,\mathbb{E}_g\bigl[\Tr_{X_i}\bigr]=N^D.
\end{equation*}
Hence, under the hypothesis of Theorem~\ref{thm:arbitrary}, $\mathcal{F}_N=N^{-D}\log\mathbb{E}[e^{-V_g}]$ converges coefficientwise to a series $\mathcal{F}$ with
\begin{equation*}
\sum_{i=1}^Ln_ig_i\,\partial_{g_i}\mathcal{F}=m_2-1,\qquad \lim_{N\to\infty}N^{s_i-D}g_i\,\mathbb{E}_g\bigl[\Tr_{X_i}\bigr]=-g_i\,\partial_{g_i}\mathcal{F}.
\end{equation*}
For a single interaction $X$ with $2n$ vertices and scaling $s$, $N^{s-D}\,\mathbb{E}_g[\Tr_X]\to(1-m_2)/(ng)$.
\end{proposition}

\begin{proof}
All identities below hold coefficient by coefficient in $g$. Under the Gaussian density proportional to $e^{-N^{D-1}T\cdot\bar T}$, integration by parts gives
\begin{equation*}
N^{D-1}\,\mathbb{E}[T_aG]=\mathbb{E}\bigl[\partial G/\partial\bar T_a\bigr]
\end{equation*}
for every polynomial $G$ in $T,\bar T$. Taking $G=\bar T_af$ and summing over the $\mathcal{K}=N^D$ indices $a$ gives
\begin{equation*}
N^{D-1}\,\mathbb{E}[Mf]=\mathcal{K}\,\mathbb{E}[f]+\mathbb{E}[\mathcal{D}f],\qquad \mathcal{D}=\sum_a\bar T_a\,\frac{\partial}{\partial\bar T_a}.
\end{equation*}
The operator $\mathcal{D}$ multiplies a polynomial of degree $k$ in $\bar T$ by $k$, and $\Tr_{X_i}$ has degree $n_i$ in $\bar T$, so $\mathcal{D}e^{-V_g}=-\sum_in_ig_iN^{s_i}\Tr_{X_i}e^{-V_g}$. Taking $f=e^{-V_g}$ and dividing by $\mathbb{E}[e^{-V_g}]$ gives the identity.

Since $g_i\partial_{g_i}\log\mathbb{E}[e^{-V_g}]=-g_iN^{s_i}\mathbb{E}_g[\Tr_{X_i}]$, the identity divided by $N^D$ reads
\begin{equation*}
\sum_{i=1}^Ln_ig_i\,\partial_{g_i}\mathcal{F}_N=N^{-1}\mathbb{E}_g[M]-1.
\end{equation*}
The operator on the left multiplies $g^\alpha$ by $\alpha\cdot\mathbf{n}=\sum_i\alpha_in_i$, which is positive for $\alpha\neq0$, and $\mathcal{F}_N$ has no constant term. Hence, for $\alpha\neq0$, $[g^\alpha]\,\mathcal{F}_N=[g^\alpha]\bigl(N^{-1}\mathbb{E}_g[M]\bigr)/(\alpha\cdot\mathbf{n})$, which converges to $[g^\alpha]\,m_2/(\alpha\cdot\mathbf{n})$ under the hypothesis, by Lemma~\ref{rem:bounded}. So $\mathcal{F}_N\to\mathcal{F}$ with $\sum_in_ig_i\partial_{g_i}\mathcal{F}=m_2-1$, and $N^{s_i-D}g_i\,\mathbb{E}_g[\Tr_{X_i}]=-g_i\partial_{g_i}\mathcal{F}_N\to-g_i\partial_{g_i}\mathcal{F}$. For $L=1$ the first relation is $ng\mathcal{F}'=m_2-1$, which gives the last claim.
\end{proof}

The identity is the simplest Schwinger--Dyson equation of tensor models~\cite{gurau2012schwinger}. It appears in~\cite[(10)]{bonzom2013revisiting} for melonic interactions at the canonical scaling and in~\cite[(3.20)]{uncolouring} for arbitrary interactions with degree-dependent scalings; the proof above does not use the scaling.

\begin{corollary}[Free-energy criterion]\label{cor:freeenergy}
Let $\mathcal{F}_N=N^{-D}\log\mathbb{E}[e^{-V_g}]$. For every $\alpha\neq0$, exactly,
\begin{equation*}
[g^\alpha]\,N^{-1}\mathbb{E}_g[M]=(\alpha\cdot\mathbf{n})\,[g^\alpha]\,\mathcal{F}_N,\qquad [g^\alpha]\,\mathcal{F}_N=\frac{(-1)^{|\alpha|}}{\alpha!}\sum_{\pi}N^{\alpha\cdot s-D+F(\pi)-(D-1)m},
\end{equation*}
the sum running over the block-connected pairings $\pi$ of the blocks $X^{(\alpha)}$, with $m$ black vertices in total. Hence the following are equivalent, for each fixed $\alpha$ and hence also for all $\alpha$ at once:
\begin{enumerate}[(i)]
\item $[g^{\alpha'}]\,N^{-1}\mathbb{E}_g[M]$ is bounded as $N\to\infty$ for every $\alpha'\leq\alpha$;
\item $[g^{\alpha'}]\,\mathcal{F}_N$ is bounded as $N\to\infty$ for every $\alpha'\leq\alpha$;
\item every connected vacuum Feynman graph with at most $\alpha_i$ copies of $X_i$ has amplitude $O(N^D)$, that is, $\alpha'\cdot s+F(\pi)-(D-1)m\leq D$ for every block-connected pairing $\pi$ of $X^{(\alpha')}$, $0\neq\alpha'\leq\alpha$.
\end{enumerate}
In particular, the hypotheses of Theorems~\ref{thm:arbitrary} and~\ref{thm:factor} hold whenever the free energy of the model is known to have a $1/N$ expansion with leading order $N^D$.
\end{corollary}

\begin{proof}
The first identity is the coefficient of $g^\alpha$ in $\sum_in_ig_i\partial_{g_i}\mathcal{F}_N=N^{-1}\mathbb{E}_g[M]-1$ from the proof of Proposition~\ref{prop:ward}, since $\sum_in_ig_i\partial_{g_i}$ multiplies $g^\alpha$ by $\alpha\cdot\mathbf{n}>0$. For the second, expanding $\log\mathbb{E}[e^{-V_g}]$ in $g$ gives $[g^\alpha]\log\mathbb{E}[e^{-V_g}]=\frac{(-1)^{|\alpha|}}{\alpha!}N^{\alpha\cdot s}\kappa(\Tr_{X^{(\alpha)}})$, the joint cumulant of the $|\alpha|$ traces. Lemma~\ref{lem:diagram} with $q=|\alpha|-1$ expresses it through block-connected pairings. This gives (i)$\Leftrightarrow$(ii). For a fixed $\alpha'$ all terms of the sum carry the same sign $(-1)^{|\alpha'|}$, so there is no cancellation between them, and the sum is bounded if and only if every exponent is at most $0$, which is (iii).
\end{proof}

The equivalence (ii)$\Leftrightarrow$(iii) is the standard criterion, cf.~\cite[\S II.A--B]{bonzom2015enhancing}; combined with (i)$\Leftrightarrow$(ii), it turns the hypothesis of Theorem~\ref{thm:arbitrary} into the usual statement that a model has a large-$N$ expansion: a bound on the connected vacuum graphs. Such bounds are what is proved, for instance, for the canonical scaling $s_i=D-1$ with arbitrary connected $X_i$, where the degree gives $F(\pi)-(D-1)m+(D-1)|\alpha|\leq D$ \cite{gurau2011expansion,gurau2012complete,uncolouring,gurau2017random}, and for the enhanced and tetrahedral models (Proposition~\ref{prop:enhanced} and Proposition~\ref{prop:tetra}). For the canonical scaling it is known that the non-melonic interactions do not contribute at leading order \cite{uncolouring,gurau2017random}, so that $m_2$ is the series $x(g)$ of the melonic interactions alone; we do not prove this here.

\section{Applications}\label{sec:applications}

\begin{corollary}[Melonic models]\label{cor:overlap}
Let $X_1,\dots,X_L$ be connected melonic and $s_i=D-1$. Then the hypothesis of Theorem~\ref{thm:arbitrary} holds, $m_2=x(g)$ with $p_i=n_i$ in \eqref{eq:twopoint}, $\bar\tau_{e_i}=1$ for every $i$, and $\bar\tau_\beta=0$ whenever $|\beta|\geq2$.
\end{corollary}

\begin{proof}
Corollary~\ref{cor:moments} with $B=\ominus$, for which $n=1$ and $\psi_0=\ell=1$, gives $[g^\alpha]\,N^{-1}\mathbb{E}_g[M]\to[g^\alpha]\,x(g)$, which is the hypothesis, with $m_2=x$. Moreover $\tau_{e_i}=N^{n_i-1}\mathbb{E}[Z_i]=N^{-1}\mathbb{E}[\Tr_{X_i}]/\mathbb{E}[\xi^{n_i}]\to1$ by Lemma~\ref{lem:radial} and (M2). Comparing \eqref{eq:m2} with \eqref{eq:twopoint} gives $\sum_{|\beta|\geq2}d_\beta\bar\tau_\beta g^\beta x^{d_\beta}=0$, and since $x(0)=1$, the terms of lowest degree show that $\bar\tau_\beta=0$ whenever $|\beta|\geq2$.
\end{proof}

That is, at large $N$ the joint cumulants on the sphere of melonic interactions vanish beyond first order. In melonic models, Theorem~\ref{thm:arbitrary} adds to Corollary~\ref{cor:moments} the factorization of melonic multi-traces. On melonic observables in melonic models the two theorems overlap: for connected melonic $B$, Corollary~\ref{cor:moments} and Theorem~\ref{thm:arbitrary} both give the limit $x(g)^n$.

\begin{proposition}[Enhanced models]\label{prop:enhanced}
Consider the generic quartic model in rank four with maximally enhanced non-melonic interactions \cite[(11)]{bonzom2015enhancing}, or a model of the family \cite[(24)]{bonzom2015enhancing}, whose interactions are arbitrary melonic graphs and trees of necklaces of colour type $12$, in both cases with the scalings of \cite{bonzom2015enhancing}. Then the hypothesis of Theorem~\ref{thm:arbitrary} holds, and coefficientwise
\begin{equation}\label{eq:enhanced}
N^{-r}\,\mathbb{E}_g\Bigl[\prod_{j=1}^r\Tr_{C_j}\Bigr]\longrightarrow m_2^{p_1+\dots+p_r}
\end{equation}
for all connected melonic $C_1,\dots,C_r$ with $2p_1,\dots,2p_r$ vertices.
\end{proposition}

\begin{proof}
The Gaussian measure of \cite[(6)]{bonzom2015enhancing} is ours with $D=4$. For the first model, in the notation of \eqref{eq:Vs}, the four quartic melonic bubbles have $s_i=3$ and the three necklaces, of colour types $12$, $13$ and $14$, have $s_i=4$. By $U(N)^4$-invariance, $N^{-1}\mathbb{E}_g[T\cdot\bar T]=N^3G_2$, where $G_2$ is the two-point function of \cite[(14)]{bonzom2015enhancing}. Each $[g^\alpha]\,N^3G_2$ is a finite sum over connected two-point maps $\Gamma$ of the intermediate-field representation, each contributing a constant times $N^{-\Omega(\Gamma)}$, and $\Omega(\Gamma)\geq0$ for every connected two-point map \cite[Thm.~1]{bonzom2015enhancing}. The hypothesis therefore holds. For the family \cite[(24)]{bonzom2015enhancing}, each connected vacuum graph is mapped to one of the restricted quartic model with the same degree \cite[Lemma~3]{bonzom2015enhancing}, and hence, by \cite[Thm.~1]{bonzom2015enhancing}, every connected vacuum graph is $O(N^4)$. This is condition~(iii) of Corollary~\ref{cor:freeenergy}, which gives the hypothesis. In both cases Theorem~\ref{thm:arbitrary} gives \eqref{eq:enhanced}.
\end{proof}

The large-$N$ limit of the first model is not melonic: its leading-order maps are trees of melonic edges joining planar components of necklace edges, glued in a cactus-like fashion \cite[\S III.D]{bonzom2015enhancing}, and its Schwinger--Dyson equations were left unsolved in \cite{bonzom2015enhancing}. By \eqref{eq:enhanced}, its whole melonic sector, products included, is nevertheless determined by the single series $m_2$.

For the family \cite[(24)]{bonzom2015enhancing}, Bonzom, Delepouve and Rivasseau proved that the expectation of a single tree of necklaces factorizes over its necklaces \cite[Thm.~3]{bonzom2015enhancing}. Since melonic graphs are the trees of necklaces of size one~\cite[\S IV.A]{bonzom2015enhancing}, this contains the case $r=1$ of \eqref{eq:enhanced}, with $m_2$ equal to their $C_1$ \cite[Thm.~3]{bonzom2015enhancing}. Theorem~\ref{thm:arbitrary} adds the products of melonic traces and the full model with necklaces of all three colour types. Non-melonic observables are not covered by Theorem~\ref{thm:arbitrary}: their products need not factorize even for Gaussian tensors~\cite{gurau2025nonfactor}.

\section{Discussion}\label{sec:discussion}

Theorem~\ref{thm:main} settles \cite[Conj.~I]{BTM} for arbitrary observables and in every rank $D\geq 3$. It gives a closed form for the correlators of \cite[Conj.~II]{BTM}, which proves the melonic-polynomial conjecture \cite[Thm.-Conj.~4.4]{TwofoldUniversality}. Theorem~\ref{thm:arbitrary} shows that melonic observables factorize, coefficientwise, in every model whose two-point function has bounded Taylor coefficients, so that the whole melonic sector of such a model is determined by the single series $m_2$. We close with some open questions.

\emph{Non-melonic observables in non-melonic models.} In melonic models, Corollary~\ref{cor:moments} gives the leading term of $\mathbb{E}_g[\Tr_B]$ for every connected $B$. Remark~\ref{rem:nonmelonic} shows that Theorem~\ref{thm:main} fails for non-melonic interactions, and Theorem~\ref{thm:arbitrary} covers melonic observables only. For these, the single-trace law and the factorization of products are equivalent. Indeed, $\prod_j\Tr_{C_j}$ is the trace of $C_1\sqcup\dots\sqcup C_r$, so the melonic reduction gives each Taylor coefficient of $N^{-r}\mathbb{E}_g[\prod_j\Tr_{C_j}]$ the leading term of the corresponding coefficient of $\mathbb{E}_g[\xi^{p_1+\dots+p_r}]$; both statements say that the limiting moments of $\xi$ are the powers of its limiting mean $m_2$ (Theorem~\ref{thm:factor}(ii)).

For a non-melonic $B$ with $2n$ vertices, the analogous law, with leading term $\ell(B)\,m_2^nN^{\psi_0(B)}$, can fail. Take $D=3$ and $V_g=\frac g3N^{3}\Tr_{K_{3,3}}$, one step above the canonical scaling. The series computed with \texttt{feyntensor} in~\cite[\S5, after (5.10c)]{BTM} gives
\begin{equation*}
m_2=1-3g+28g^2-351g^3+O(g^4).
\end{equation*}
Here the exact identity of Proposition~\ref{prop:ward} reads
\begin{equation*}
N^{-1}\mathbb{E}_g[M]+g\,\mathbb{E}_g[\Tr_{K_{3,3}}]=1,
\end{equation*}
so the coefficient of $N^0$ in $\mathbb{E}_g[\Tr_{K_{3,3}}]$ is
\begin{equation*}
\frac{1-m_2}{g}=3-28g+351g^2+O(g^3),\qquad\text{whereas}\qquad \ell(K_{3,3})\,m_2^3=3-27g+333g^2+O(g^3).
\end{equation*}
What replaces this law is open. Products of non-melonic traces need not factorize even for Gaussian tensors \cite{gurau2025nonfactor,CCL26}; whether families that factorize in the Gaussian case, such as those with tree-like dominant pairings \cite{CCL26}, still factorize in interacting models is also open.

\emph{The hypothesis of Theorem~\ref{thm:arbitrary}.} Corollary~\ref{cor:freeenergy} reduces it to a bound on connected vacuum graphs. Outside melonic models (Corollary~\ref{cor:overlap}) that bound is verified here only through results proved elsewhere: the degree bound at the canonical scaling~\cite{gurau2011expansion,gurau2012complete,uncolouring,gurau2017random}, \cite[Thm.~1]{bonzom2015enhancing} for the enhanced models, and \cite[Prop.~2]{carrozza2016on} for the tetrahedral model. No criterion is known that can be checked directly on the $(X_i,s_i)$, or that identifies the largest admissible scalings~\cite{RedundantMoments}.

\emph{Fixed coupling.} All results here are coefficientwise. A fixed-coupling version of Theorem~\ref{thm:arbitrary}, even restricted to melonic models, is open.

\addtocontents{toc}{\protect\setcounter{tocdepth}{0}}
\section*{Acknowledgements}
This work was supported by a McGill Graduate Excellence Award and by the NSERC Discovery Grant of Elliot Paquette.

\section*{AI Declaration}
\addtocontents{toc}{\protect\setcounter{tocdepth}{1}}
During the preparation of this work the author used Claude Opus 5.5 (Anthropic). It produced proofs of Conjecture~\ref{conj:I} for specific non-melonic graphs; the author checked and streamlined these proofs, which helped in proving Conjecture~\ref{conj:I}. The author then generalized the result to Theorem~\ref{thm:main}. The model helped to extend the results to real tensors with $O(N)^D$ symmetry (\hyperref[app:real]{Appendix}). It was further used for literature review, figure creation, rough drafting and editing, and to referee sections of the paper over several rounds of review and revision. The author takes full responsibility for the content of this paper.

\appendix
\makeatletter
\let\@origseccntformat\@seccntformat
\let\@origtocwrite\@tocwrite
\let\sectionname\relax
\def\@seccntformat#1{\appendixname\protect\textup{\protect\@secnumpunct}}
\def\@tocwrite#1{\@tocwriteb\tocappendixsingle{#1}}
\section{Real tensors}\label{app:real}
\let\@seccntformat\@origseccntformat
\let\@tocwrite\@origtocwrite
\let\sectionname\appendixname
\makeatother
This appendix extends Theorems~\ref{thm:main} and~\ref{thm:arbitrary} to real tensors with $O(N)^D$ symmetry \cite{carrozza2016on,klebanov2017uncolored}. Their invariants are indexed by $D$-coloured graphs that need not be bipartite, and their Wick pairings are arbitrary perfect matchings. Real rank-three tensors in irreducible representations of $O(N)$, symmetric traceless, antisymmetric or of mixed symmetry, also have a melonic $1/N$ expansion with tetrahedral interaction~\cite{benedetti2019symmetric,carrozza2018irreducible}; we do not treat them here. Real tensor models with $O(N)^D$ symmetry are related to symplectic ones by an $N\to-N$ duality~\cite{gurau2025duality,keppler2023duality}.

\subsection{Setting}\label{ssec:real}
Let $T\in\mathbb{R}^{N^D}$ have the Gaussian density proportional to $\exp(-\tfrac12N^{D-1}\,T\cdot T)$, so that $\mathbb{E}[T_aT_b]=N^{-(D-1)}\delta_{ab}$. For real tensors, a \emph{$D$-coloured graph} is any finite $D$-regular graph with a proper edge colouring by $[D]$, not necessarily bipartite, such as the tetrahedron $K_4$ in rank $3$, in which every two of four vertices are joined by one edge (Fig.~\ref{fig:tetra}). We write $\eta_c$ for the perfect matching formed by its colour-$c$ edges. Attaching to every vertex $v$ a factor $T_{a_v^1\cdots a_v^D}$ and identifying $a_v^c=a_w^c$ along every colour-$c$ edge $\{v,w\}$ defines the $O(N)^D$-invariant $\Tr_B(T)$, homogeneous of degree $2n$ if $B$ has $2n$ vertices. For bipartite $B$ it is the invariant of Section~\ref{ssec:coloured} with $\bar T$ replaced by $T$. Since melonic graphs are bipartite, melonic interactions, (M1) and slots are as before.

Given blocks $Y_0,\dots,Y_q$, now arbitrary $D$-coloured graphs with $2m$ vertices in total, a \emph{Wick pairing} $\pi$ is a perfect matching of all their vertices, with $m$ propagators. The faces of colour $(0,c)$ are the cycles of the $2$-regular graph $\pi\cup\eta_c$, and $F(\pi)$ is their total number over $c\in[D]$. Block-connectivity, $\chi$, $\Psi$ and $\Lambda$ are defined as in Sections~\ref{ssec:gaussian} and~\ref{ssec:blocks}.

\begin{lemma}[Real Wick and diagram formulas]\label{lem:wickO}
$\mathbb{E}\bigl[\prod_i\Tr_{Y_i}\bigr]=\sum_\pi N^{F(\pi)-(D-1)m}$, the sum running over all perfect matchings, and $\kappa(\Tr_{Y_0},\dots,\Tr_{Y_q})$ is the same sum restricted to block-connected $\pi$. In particular \eqref{eq:diagchi} holds for real tensors.
\end{lemma}

\begin{proof}
As in Lemma~\ref{lem:wick}: the only non-zero joint cumulant of the entries is $\kappa(T_a,T_b)=N^{-(D-1)}\delta_{ab}$, now between any two entries. A propagator identifies all $D$ indices of its two ends, so each cycle of $\pi\cup\eta_c$ carries one free index.
\end{proof}

We define $\mathcal{I}^O_N[B;X_1,\dots,X_q]$ by \eqref{eq:IN} with the real cumulant and $\psi(\pi)$ by \eqref{eq:psi}, with $m$ the number of propagators. For connected $B$ with $2n$ vertices let $\psi_0^O(B)$ be the largest $F(\pi)-(D-1)n$ over the pairings of $B$ alone and $\ell^O(B)$ the number of pairings attaining it, so that $N^{-1}\mathbb{E}[\Tr_B]=\ell^O(B)N^{\psi^O_0(B)-1}+O(N^{\psi^O_0(B)-2})$.
Moreover, \eqref{eq:INchi} holds for $\mathcal{I}^O_N$ as well.

\begin{theorem}\label{thm:mainO}
Let $D\geq 3$, let $B$ be a connected $D$-coloured graph with $2n$ vertices, and let $X_1,\dots,X_q$ be connected melonic graphs with $2p_1,\dots,2p_q$ vertices. Then
\begin{equation*}
\mathcal{I}^O_N[B;X_1,\dots,X_q]=2^q\,\ell^O(B)\,\mathrm{Mel}_{q+1}(n,p_1,\dots,p_q)\,N^{\psi^O_0(B)-1}+O\bigl(N^{\psi^O_0(B)-2}\bigr).
\end{equation*}
\end{theorem}

\subsection{Melonic reduction}\label{ssec:reductionO}
For real tensors, dipole splitting acquires a third kind of pairing, which is again suppressed. Let the block $Y_i$ contain a connected melonic component $Y$ with at least four vertices, and let $\delta=\{k,w\}$ be a $(D-1)$-dipole of $Y$ missing the colour $c$ such that $Y\setminus\delta$ is melonic, as in (M1). Put $w'=\eta_c(k)$ and $k'=\eta_c(w)$. Define $Y_i^\flat$ by replacing the colour-$c$ edges $\{k,w'\},\{w,k'\}$ by $\{k,w\},\{k',w'\}$, so that $Y$ becomes $(Y\setminus\delta)\sqcup\ominus$, and $Y_i''$ by replacing $Y$ with $Y\setminus\delta$. As before, $\chi^\flat$, $\chi''$ denote the exponents \eqref{eq:chi} computed with them.

\begin{lemma}[Splitting]\label{lem:splitO}
For every pairing $\pi$, block-connectivity is the same for $Y_i$ and $Y_i^\flat$, and $\chi^\flat(\pi)=\chi(\pi)+\Delta-1$, where $\Delta\in\{+1,0,-1\}$ is the change in the number of faces of colour $(0,c)$. We call $\pi$ \emph{good}, \emph{twisted} or \emph{bad} according as $\Delta=+1,0,-1$.
\end{lemma}

\begin{proof}
The blocks are unchanged. Only colour-$c$ edges change: $\pi\cup\eta^\flat_c$ arises from $\pi\cup\eta_c$ by removing two edges and rejoining their four ends, which splits, keeps or merges the faces through them. Hence $\Delta\in\{+1,0,-1\}$ and no other face changes. The extra component of $Y_i^\flat$ contributes $-1$ to \eqref{eq:chi}.
\end{proof}

\begin{lemma}[Pairings that are not good]\label{lem:notgoodO}
Let $\pi$ be a pairing with $\pi(k)\neq w$. Put $y=\pi(w)$, $u=\pi(k)$, as in Lemma~\ref{lem:bad}, and let $\pi'$ be obtained from $\pi$ by replacing the propagators $\{w,y\},\{k,u\}$ with $\{y,u\}$, a perfect matching of the vertices other than $k,w$. Then
\begin{equation*}
\chi(\pi)\leq\chi''(\pi')-(D-2).
\end{equation*}
Every pairing that is not good satisfies $\pi(k)\neq w$, and if it is block-connected, then $\pi'$ is block-connected for the blocks with $Y_i''$ in place of $Y_i$.
\end{lemma}

\begin{proof}
If $\pi(k)=w$, the face of colour $(0,c)$ through $k$ contains $w',k,w,k'$ in this order, and the rewiring closes off the face $\{k,w\}$; thus $\Delta=+1$. Hence a pairing that is not good has $\pi(k)\neq w$. Now let $\pi(k)\neq w$; then $y,u\notin\{k,w\}$ and $y\neq u$.

\emph{Faces.} As in Lemma~\ref{lem:bad}, we compare $\pi$ on $Y_i^\flat$ with $\pi'$ on $Y_i''$. In $Y_i^\flat$ the pair $\{k,w\}$ is a melon, so for every colour $e$ the face of colour $(0,e)$ through $k$ contains the path $y,w,k,u$, which $\pi'$ shortens to the propagator $\{y,u\}$. No face disappears, since the shortened face still contains $\{y,u\}$, and the other faces are unchanged; hence $F''(\pi')=F^\flat(\pi)$. Replacing two propagators by one removes one propagator and one component of $Y_i^\flat$, which gives $\chi''(\pi')=\chi^\flat(\pi)+D$, as in Lemma~\ref{lem:bad}. By Lemma~\ref{lem:splitO}, $\chi(\pi)=\chi^\flat(\pi)+1-\Delta$ with $\Delta\geq-1$, so
\begin{equation*}
\chi(\pi)=\chi''(\pi')-(D-1)-\Delta\leq\chi''(\pi')-(D-2).
\end{equation*}

\emph{Connectivity.} Let $\pi$ be block-connected and suppose the block graph of $\pi'$ is disconnected, with $A$ the component containing $Y_i$. The blocks of $y$ and $u$ lie outside $A$, since otherwise no propagator of $\pi$ would leave $A$. Hence $\{k,u\}$ and $\{w,y\}$ are the only propagators of $\pi$ leaving $A$. The face of colour $(0,c)$ through $k$ then reads $k,u,\dots,y,w,k',\dots,w',k$, and the rewiring splits it into $k,u,\dots,y,w,k$ and $k',\dots,w',k'$. Thus $\pi$ is good.
\end{proof}

\begin{proposition}\label{prop:melredO}
Proposition~\ref{prop:melred} and Corollary~\ref{cor:reduction} hold for real tensors, for blocks that are arbitrary $D$-coloured graphs, not necessarily bipartite.
\end{proposition}

\begin{proof}
Monotonicity is proved as before. In the induction, Lemmas~\ref{lem:splitO} and~\ref{lem:notgoodO} replace Lemmas~\ref{lem:split} and~\ref{lem:bad}: a twisted or bad pairing has $\chi(\pi)\leq\chi''(\pi')-(D-2)\leq\Psi_{p-1}-(D-2)<\Psi_p$ and $\chi^\flat(\pi)\leq\chi(\pi)-1<\Psi_p$, while a good pairing has $\chi(\pi)=\chi^\flat(\pi)$. The pairings attaining $\Psi_p$ are therefore the same for $Y_i$ and $Y_i^\flat$, and the rest of the proof is unchanged.
\end{proof}

\begin{lemma}\label{lem:melonicO}
If $B$ is connected and melonic, then $\psi^O_0(B)=\ell^O(B)=1$.
\end{lemma}

\begin{proof}
We induct on the number of vertices. The melon has one pairing, with $F=D$ and hence $\psi=1$. Otherwise we choose a dipole $\delta=\{k,w\}$ with $B\setminus\delta$ melonic, by (M1). Pairings with $\pi(k)=w$ correspond bijectively to pairings of $B\setminus\delta$, with $D-1$ more faces and one more propagator, hence the same exponent. Every other pairing has $\pi(k)\neq w$, hence exponent at most $1-(D-2)<1$ by Lemma~\ref{lem:notgoodO} and the induction hypothesis.
\end{proof}

\subsection{The radial variable}\label{ssec:radialO}
We set $M=T\cdot T$ and $\xi=M/N$. Since $N^{D-1}M$ is a sum of $\mathcal{K}=N^D$ squares of independent standard Gaussians, $\xi\sim\mathrm{Gamma}(\mathcal{K}',\ \text{rate }\mathcal{K}')$ with $\mathcal{K}'=\mathcal{K}/2$, and the direction $T/\sqrt M$ is uniform on the sphere and independent of $M$. As $\Tr_B$ is homogeneous of degree $2n$, we have $\Tr_B=M^nZ_B$ with $Z_B$ independent of $M$. Again $\Tr_{\mathcal{M}_p}=M^p$. The proof of Lemma~\ref{lem:radial} and the computation leading to \eqref{eq:radial} use only these facts and therefore give
\begin{equation}\label{eq:radialO}
\mathcal{I}^O_N\bigl[B;\mathcal{M}_{p_1},\dots,\mathcal{M}_{p_q}\bigr]=N^{-1}\mathbb{E}[\Tr_B]\cdot\frac{\mathcal{K}^q\,\kappa(\xi^n;\xi^{p_1},\dots,\xi^{p_q})}{\mathbb{E}[\xi^n]}.
\end{equation}
Lemmas~\ref{lem:loop} and~\ref{lem:poly} and Proposition~\ref{prop:gammalimit} use the law of $\xi$ only through \eqref{eq:gamma}, which holds for every real $\mathcal{K}>0$. With $\mathcal{K}'$ in place of $\mathcal{K}$ and $\mathcal{K}^q=2^q\mathcal{K}'^q$, the second factor in \eqref{eq:radialO} is therefore
\begin{equation}\label{eq:HO}
2^q\,\mathrm{Mel}_{q+1}(n,p_1,\dots,p_q)+O(N^{-D}).
\end{equation}
Likewise, in Section~\ref{ssec:interacting} for real tensors, $\xi\sim\mathrm{Gamma}(\mathcal{K}',\text{rate }\mathcal{K}')$ and $\mathcal{K}$ is replaced by $\mathcal{K}'$ in \eqref{eq:tau}, \eqref{eq:W} and \eqref{eq:recs}, since Lemma~\ref{lem:loop} uses only \eqref{eq:gamma}.

\subsection{Proofs}\label{ssec:proofsO}
\begin{proof}[Proof of Theorem~\ref{thm:mainO}]
By Proposition~\ref{prop:melredO} and \eqref{eq:INchi}, $\mathcal{I}^O_N[B;X_1,\dots,X_q]$ and $\mathcal{I}^O_N[B;\mathcal{M}_{p_1},\dots,\mathcal{M}_{p_q}]$ have the same top exponent and coefficient. By \eqref{eq:radialO} and \eqref{eq:HO} the latter has top term\begin{equation*}2^q\,\ell^O(B)\,\mathrm{Mel}_{q+1}(n,p_1,\dots,p_q)\,N^{\psi^O_0(B)-1},\end{equation*}which is non-zero, and the exponents are integers.
\end{proof}

\begin{corollary}\label{cor:melonicO}\label{cor:momentsO}
Corollaries~\ref{cor:melonic} and~\ref{cor:moments} hold under the real Gaussian measure, with $(\psi^O_0,\ell^O)$ in place of $(\psi_0,\ell)$, with an extra factor $2^q$ in the limit of Corollary~\ref{cor:melonic}, and with the potential $V^O_g=\tfrac12N^{D-1}\sum_{i=1}^Lg_i\Tr_{X_i}$. In particular, real and complex melonic models with the same couplings have the same series $x(g)$.
\end{corollary}

\begin{proof}
As before, with Lemma~\ref{lem:melonicO} for $C_0$; in the second statement the factor $2^{-|\alpha|}$ from $V^O_g$ cancels the factor $2^{|\alpha|}$ of Theorem~\ref{thm:mainO}.
\end{proof}

\begin{figure}[t]
\centering
\begin{subfigure}[b]{0.3\textwidth}
\centering
\begin{tikzpicture}[x=0.75cm, y=0.75cm]
\coordinate (v1) at (0,0);
\coordinate (v2) at (90:1.6); \coordinate (v3) at (210:1.6); \coordinate (v4) at (330:1.6);
\draw[lbc1] (v1) -- (v2); \draw[lbc1] (v3) -- (v4);
\draw[lbc2] (v1) -- (v3); \draw[lbc2] (v2) -- (v4);
\draw[lbc3] (v1) -- (v4); \draw[lbc3] (v2) -- (v3);
\foreach \i in {1,...,4} \node[lbwv] at (v\i) {};
\node[lblab, anchor=north] at ([yshift=-3pt]v1) {$1$};
\node[lblab, anchor=south] at ([yshift=2pt]v2) {$2$};
\node[lblab, anchor=north east] at ([xshift=-1pt,yshift=-1pt]v3) {$3$};
\node[lblab, anchor=north west] at ([xshift=1pt,yshift=-1pt]v4) {$4$};
\end{tikzpicture}
\caption{$K_4$}
\end{subfigure}%
\begin{subfigure}[b]{0.66\textwidth}
\centering
\begin{tikzpicture}[x=0.42cm, y=0.42cm]
\newcommand{\tetpanel}[6]{%
  \begin{scope}[shift={(#1,#2)}]
    \coordinate (t1) at (0,0);
    \coordinate (t2) at (90:1.6); \coordinate (t3) at (210:1.6); \coordinate (t4) at (330:1.6);
    \foreach \a/\b in {#4} \draw[#3] (t\a) -- (t\b);
    \foreach \a/\b in {#5} \draw[lbprop] (t\a) to[bend right=30] (t\b);
    \foreach \i in {1,...,4} \node[lbwv] at (t\i) {};
    \node[lbelab] at (0,-1.75) {#6};
  \end{scope}}
\foreach \c/\x in {1/0,2/5,3/10} \node[lblab] at (\x,2.8) {colour $(0,\c)$};
\tetpanel{0}{0}{lbc1}{1/2,3/4}{1/2,3/4}{$2$}
\tetpanel{5}{0}{lbc2}{1/3,2/4}{1/2,3/4}{$1$}
\tetpanel{10}{0}{lbc3}{1/4,2/3}{1/2,3/4}{$1$}
\tetpanel{0}{-4.6}{lbc1}{1/2,3/4}{1/3,2/4}{$1$}
\tetpanel{5}{-4.6}{lbc2}{1/3,2/4}{1/3,2/4}{$2$}
\tetpanel{10}{-4.6}{lbc3}{1/4,2/3}{1/3,2/4}{$1$}
\tetpanel{0}{-9.2}{lbc1}{1/2,3/4}{1/4,2/3}{$1$}
\tetpanel{5}{-9.2}{lbc2}{1/3,2/4}{1/4,2/3}{$1$}
\tetpanel{10}{-9.2}{lbc3}{1/4,2/3}{1/4,2/3}{$2$}
\foreach \p/\y in {1/0.3, 2/-4.3, 3/-8.9} {
  \node[lblab, anchor=east] at (-2.3,\y) {$\pi=\eta_{\p}$};
  \node[lblab, anchor=west] at (12.3,\y) {$F=4$};
}
\end{tikzpicture}
\caption{faces of each colour, and their number}
\end{subfigure}
\captionsetup{font=footnotesize}
\caption{The tetrahedron $K_4$ in rank $3$; colour $1$ is red, $2$ blue and $3$ green, and every vertex carries $T$. (a) $K_4$, with each colour class $\eta_c$ a perfect matching. (b) Its three Wick pairings $\pi=\eta_1,\eta_2,\eta_3$, drawn dashed together with the colour-$c$ edges, for each $c$. The number below each panel is its number of cycles, and $F$ is their total.}\label{fig:tetra}
\end{figure}
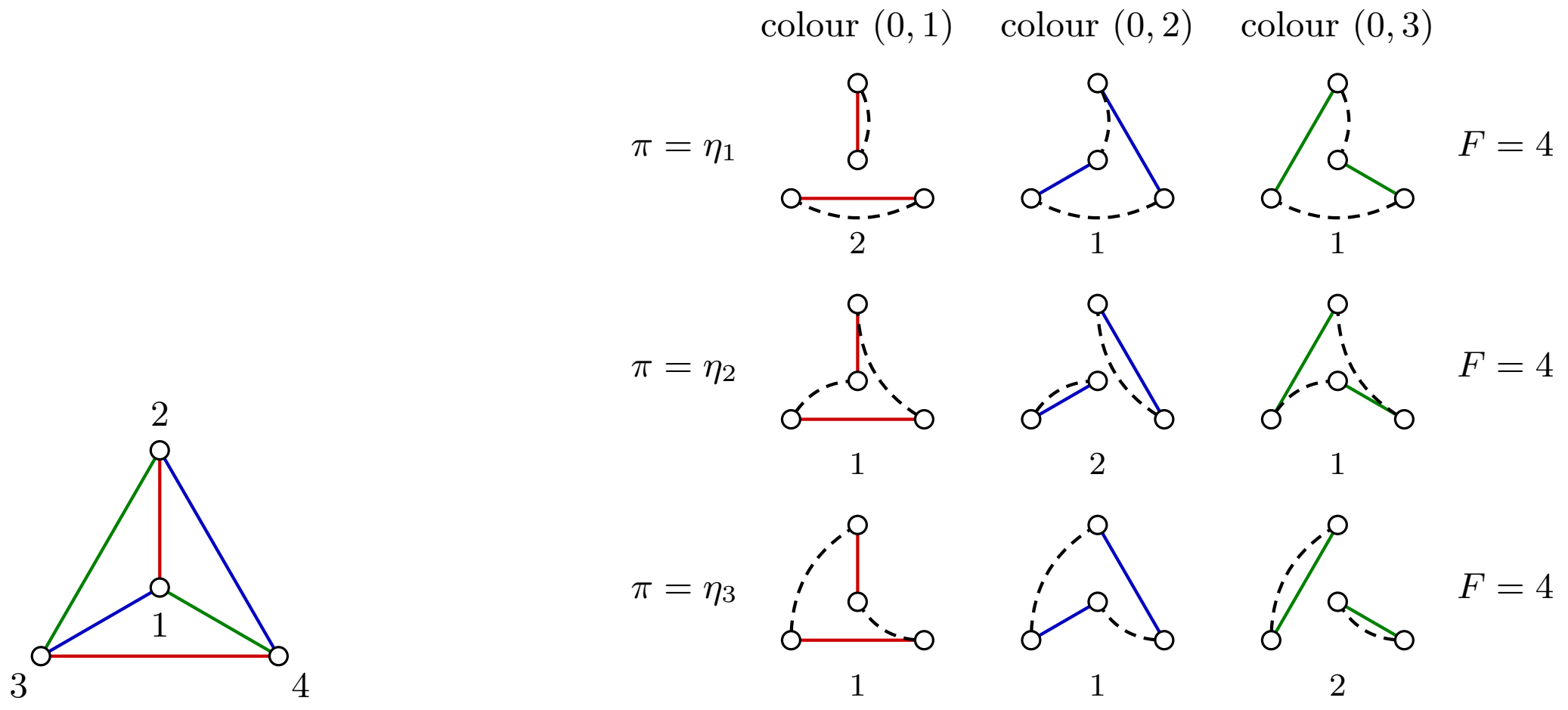

\begin{remark}[Real and complex exponents]\label{rem:realcomplex}
Every black--white pairing of a bipartite $B$ is a real pairing with the same faces. Hence $\psi^O_0(B)\geq\psi_0(B)$, and $\ell^O(B)\geq\ell(B)$ when equality holds. Theorem~\ref{thm:mainO} also applies to non-bipartite observables, which have no complex counterpart: for $K_4$, all three pairings have exponent $0$ (Fig.~\ref{fig:tetra}). Each pairing is a colour class $\eta_c$, and it has two faces of colour $(0,c)$ and one face of each other colour; thus $F=4=(D-1)m$. Hence $\mathbb{E}[\Tr_{K_4}]=3$ and $\mathcal{I}^O_N[K_4;X_1,\dots,X_q]=3\cdot2^q\,\mathrm{Mel}_{q+1}(2,p_1,\dots,p_q)\,N^{-1}+O(N^{-2})$ for all connected melonic $X_1,\dots,X_q$.
\end{remark}

\subsection{Factorization of melonic observables}\label{ssec:arbreal}
Theorems~\ref{thm:arbitrary} and~\ref{thm:factor} and Proposition~\ref{prop:ward} hold for real tensors, with arbitrary connected $D$-coloured interactions $X_i$, not necessarily bipartite, $X_i$ having $2n_i$ vertices, and with $M=T\cdot T$. The proofs are unchanged, with $\mathcal{K}'=\mathcal{K}/2$ in place of $\mathcal{K}$ (Section~\ref{ssec:radialO}) and Proposition~\ref{prop:melredO} in place of Proposition~\ref{prop:melred}, except for the Ward identity. There integration by parts reads $\mathbb{E}[T_aG]=N^{-(D-1)}\mathbb{E}[\partial_aG]$ and $T\cdot\nabla\Tr_{X_i}=2n_i\Tr_{X_i}$. The Ward identity therefore becomes
\begin{equation*}
N^{D-1}\,\mathbb{E}_g[M]+\sum_{i=1}^L2n_ig_iN^{s_i}\,\mathbb{E}_g\bigl[\Tr_{X_i}\bigr]=N^D,\qquad \sum_{i=1}^L2n_ig_i\,\partial_{g_i}\mathcal{F}=m_2-1.
\end{equation*}
Corollary~\ref{cor:freeenergy} holds with $2\alpha\cdot\mathbf{n}$ in place of $\alpha\cdot\mathbf{n}$ and the real pairings of Lemma~\ref{lem:wickO}, whose weights are again positive. The potential \eqref{eq:Vs} for melonic interactions with $s_i=D-1$ is $V^O_{2g}$ in the notation of Corollary~\ref{cor:momentsO}. For real tensors, Corollary~\ref{cor:overlap} therefore holds with $m_2=x(2g)$ and $\bar\tau_{e_i}=2$, the extra factor coming from $\mathcal{K}'=\mathcal{K}/2$ in \eqref{eq:tau}.

\subsection{The tetrahedral model}\label{ssec:tetra}
We consider the tetrahedral model of Carrozza and Tanasa~\cite{carrozza2016on}, of which the model of Klebanov and Tarnopolsky~\cite{klebanov2017uncolored} is a fermionic quantum-mechanical analogue (see~\cite{klebanov2018tasi} for a review).

\begin{proposition}[The tetrahedral model]\label{prop:tetra}
Let $D=3$, and consider real tensors with the potential $V_g=gN^{5/2}\Tr_{K_4}$. Then the hypothesis of Theorem~\ref{thm:arbitrary} holds, and $m_2$ is the unique formal power series with $m_2(0)=1$ and
\begin{equation*}
m_2=1+16g^2m_2^4,\qquad\text{so that}\qquad [g^{2k}]\,m_2^p=16^k\,\frac{p}{4k+p}\binom{4k+p}{k},\quad [g^{2k+1}]\,m_2^p=0.
\end{equation*}
Consequently, coefficientwise, $N^{-r}\,\mathbb{E}_g\bigl[\prod_{j=1}^r\Tr_{C_j}\bigr]\to m_2^{p_1+\dots+p_r}$ for connected melonic $C_1,\dots,C_r$ with $2p_1,\dots,2p_r$ vertices, and
\begin{equation*}
N^{-1/2}\,\mathbb{E}_g\bigl[\Tr_{K_4}\bigr]\longrightarrow\frac{1-m_2}{4g}=-4g\,m_2^4.
\end{equation*}
\end{proposition}

\begin{proof}
\emph{Normalization.} Carrozza and Tanasa consider $Z_N=\int[dT]\,e^{-N^{3/2}S_N(T)}$ \cite[(2)]{carrozza2016on} with the quartic action
\begin{equation*}
S_N=\tfrac12\,T\cdot T+\tfrac{\lambda_1}{4}\Tr_{K_4}+\tfrac{\lambda_2}{12\sqrt N}\textstyle\sum_{\ell}\Tr_{P_\ell}
\end{equation*}
\cite[(21)--(22)]{carrozza2016on}, in which the tetrahedral bubble has scaling exponent $\rho=0$ and the pillows $\rho=\frac12$; the Gaussian covariance is $N^{-3/2}$. Writing their tensor as $T_{\mathrm{CT}}=N^{1/4}T$ and setting $\lambda_2=0$ gives the density proportional to $\exp\bigl(-\tfrac12N^2\,T\cdot T-\tfrac{\lambda_1}{4}N^{5/2}\Tr_{K_4}(T)\bigr)$, which is our measure with $V_g$ for $\lambda_1=4g$.

\emph{The hypothesis.} A connected vacuum graph with $V$ tetrahedral vertices has $2V$ propagators. If it has $F$ faces, its amplitude is $N^{F-3V/2}$ in both normalizations: in ours it is $N^{Vs+F-(D-1)m}$ with $s=\frac52$ and $m=2V$, and in theirs each vertex carries $N^{3/2}$ and each propagator $N^{-3/2}$. By \cite[Prop.~2]{carrozza2016on}, this amplitude is $N^{3-\omega}$ with $\omega\geq0$. This is condition~(iii) of Corollary~\ref{cor:freeenergy} in its real version (Section~\ref{ssec:arbreal}), which therefore gives the hypothesis.

\emph{The two-point function.} Put $G=N^{-1}\mathbb{E}_g[T\cdot T]$, so that $G=1$ for the Gaussian. Each coefficient of $G$ is a sum over two-point graphs, obtained by cutting a propagator of a connected vacuum graph, with the same amplitudes. Its coefficientwise limit $m_2$ is therefore the sum over the two-point graphs with $\omega=0$, which is the leading-order two-point function $G_{\mathrm{LO}}$ of \cite[\S IV]{carrozza2016on}. In the variables $g_{\mathrm{CT}}=\lambda_1^2$ and $\mu=-\lambda_2/\lambda_1^2$ of \cite[(31)]{carrozza2016on}, the two-point equation \cite[(45)]{carrozza2016on} reads $G_{\mathrm{LO}}=1+g_{\mathrm{CT}}G_{\mathrm{LO}}^2(G_{\mathrm{LO}}^2+\mu)$; at $\lambda_2=0$, $\lambda_1=4g$ it becomes $m_2=1+16g^2m_2^4$. Equivalently, in \eqref{eq:m2}, with $d_k=2k$, $\bar\tau_2=-4$ and $\bar\tau_k=0$ for all $k\neq2$, including $k=1$: at large $N$ only the second rescaled cumulant on the sphere of $Z_{K_4}$ survives. Lagrange--B\"urmann inversion in $u=16g^2$, as in Lemma~\ref{lem:lagrange}, gives the coefficients.

\emph{Consequences.} The melonic moments follow from Theorem~\ref{thm:arbitrary} for real tensors (Section~\ref{ssec:arbreal}). The real Ward identity with $L=1$ and $n_1=2$ gives $4g\,\mathcal{F}'=m_2-1$, and, as in the proof of Proposition~\ref{prop:ward}, $N^{s-D}\,\mathbb{E}_g[\Tr_{K_4}]\to-\mathcal{F}'=(1-m_2)/(4g)$ with $s-D=-1/2$.
\end{proof}

\bibliographystyle{alpha}
\bibliography{biblio_split}
\end{document}